\documentclass[graybox]{svmult}

\usepackage{mathptmx}       % selects Times Roman as basic font
\usepackage{helvet}         % selects Helvetica as sans-serif font
\usepackage{courier}        % selects Courier as typewriter font
\usepackage{type1cm}        % activate if the above 3 fonts are
\usepackage{makeidx}         % allows index generation
\usepackage{graphicx}        % standard LaTeX graphics tool
\usepackage{multicol}        % used for the two-column index
\usepackage[bottom]{footmisc}% places footnotes at page bottom

\begin{document}
 \newcommand{\be}{\begin{equation}}
\newcommand{\ee}{\end{equation}}
 \newcommand{\bal}{\begin{align}}
  \newcommand{\eal}{\end{align}}
 \newcommand{\ben}{\begin{equation*}}
\newcommand{\een}{\end{equation*}}
\newcommand{\bea}{\begin{eqnarray}}
\newcommand{\eea}{\end{eqnarray}}
\newcommand{\bean}{\begin{eqnarray*}}
\newcommand{\eean}{\end{eqnarray*}}
\newcommand{\bes}{\begin{subequations}}
\newcommand{\ees}{\end{subequations}}
\def\so{\sigma(\omega)}
\def\Re{{\rm Re}}
\def\Im{{\rm Im}}
\def\w{\omega}
\def\pa{\partial}
\def\vp{\varphi}
\def\ocal{ O}
\def\R{{\cal R}}
\def\a{\alpha}

\title*{Introduction to Holographic Superconductors}
\author{Gary T. Horowitz}
\institute{Gary T. Horowitz \at Physics Department, UCSB, Santa Barbara, CA 93106, USA, \email{gary@physics.ucsb.edu}}
\maketitle

\abstract{These lectures give an introduction to the theory of holographic superconductors. These are superconductors that have a dual gravitational description using gauge/gravity duality. After introducing a suitable gravitational theory, we discuss its properties in various regiemes: the probe limit, the effects of backreaction, the zero temperature limit, and the addition of magnetic fields. Using the gauge/gravity dictionary, these properties reproduce many of the standard features of superconductors. Some familiarity with gauge/gravity duality is assumed. A list of open problems is included at the end.}

\section{Introduction}
\label{sec:1}

The name ``holographic superconductor" suggests that one can look at a two (spatial) dimensional superconductor and see a three dimensional image. We will see that there is a class of superconductors for which this is true, but the image one ``sees" is quite striking. It involves a charged black hole with nontrivial ``hair". This remarkable connection between condensed matter and gravitational physics was discovered just a few years ago.
It grew out of the gauge/gravity duality which has emerged from string theory \cite{Maldacena:1997re,Gubser:1998bc,Witten:1998qj}. Although this duality was first formulated as a equivalence between a certain gauge theory and a theory of quantum gravity (and provided new insights into each of these theories), over the past decade it has been applied with notable success to other areas of physics as well. Many of these new applications are discussed in other lectures in this school. I will focus on  the application to superconductivity. These lectures will be heavily based on \cite{Hartnoll:2008vx,Hartnoll:2008kx,Horowitz:2008bn,Horowitz:2009ij}. For a more general discussion of applying gauge/gravity duality to condensed matter, see the excellent reviews by Hartnoll \cite{Hartnoll:2009sz}, Herzog \cite{Herzog:2009xv} and McGreevy\cite{McGreevy:2009xe}.

I will start with a brief introduction to superconductivity. In section 2, I will introduce a simple model for a holographic superconductor. Most of our discussion will be devoted to exploring the consequences of this model, beginning in section 3 with the  probe limit - a  simplification of the model which preserves most of the physics. In section 4 we will discuss the full solution with backreaction. We next examine  the ground state of the superconductor (section 5), and study its behavior when magnetic fields are added (section 6). We conclude with a brief discussion of recent developments (section 7) and a list of open problems (section 8).  

\subsection{Superconductivity}

It was noticed in the early part of the $ 20^{th}$ century that the electrical resistivity of most metals drops suddenly to zero as the temperature is lowered below a critical temperature $T_c$. These materials were called superconductors\footnote{A good general reference is \cite{Tinkham:1996}.}.
 A second independent property of these materials was the Meissner effect: A magnetic field is expelled when $T< T_c$. This is perfect diamagnetism and does not follow from the perfect conductivity (which alone would imply that a pre-existing magnetic field is trapped inside the sample). A phenomenological description of both of these properties was first given by the London brothers in 1935 with the simple equation $J_i \propto A_i$ \cite{London:1935}. Taking a time derivative yields $E_i \propto \partial J_i/\partial t$, showing that electric fields accelerate superconducting electrons rather than keeping their velocity constant as in Ohm's law with finite conductivity. Taking the curl of both sides and combining with Maxwell's equations yields $\nabla^2 B_i \propto B_i$ showing the decay of magnetic fields inside a superconductor.

In 1950, Landau and Ginzburg described superconductivity in terms of a second order phase transition whose order parameter is a complex scalar field $\varphi$ \cite{Ginzburg:1950sr}.  The density of superconducting electrons is given by $n_s = |\varphi(x)|^2$. The contribution of $\varphi$ to the free energy is assumed to take the form
\be
F = \alpha (T-T_c) |\varphi|^2 + {\beta\over 2}|\varphi|^4 + \cdots
\ee
where $\alpha$ and $\beta$ are positive constants and the dots denote gradient terms and higher powers of $\varphi$. Clearly for $T > T_c$ the minimum of the free energy is at $\varphi = 0$ while for $T<T_c$ the minimum is at a nonzero value of $\varphi$. This is just like the Higgs mechanism in particle physics, and is associated with breaking a $U(1)$ symmetry. The London equation follows from this spontaneous symmetry breaking \cite{Weinberg86}.

A more complete theory of superconductivity was given by Bardeen, Cooper and Schrieffer in 1957 and is known as BCS theory \cite{Bardeen:1957mv}. They showed that interactions with phonons can cause 
  pairs of elections with opposite spin to bind and form a charged boson called a Cooper pair.  Below a critical temperature $T_c$, there is a second order phase transition and these bosons condense.
 The DC conductivity becomes infinite producing a superconductor. The pairs are are not bound very tightly and typically have a size which is much larger than the lattice spacing.
 In the superconducting ground state, there is an energy gap $\Delta$ for charged excitations. This gap is typically related to the critical temperature by $\Delta \approx 1.7 T_c$.  The charged excitations are ``dressed electrons" called quasiparticles. The gap in the spectrum results in a gap in the (frequency dependent) optical conductivity.  If a photon of frequency $\omega$ hits the superconductor, it must produce two quasiparticles. The binding energy of the Cooper pair is very small, but the energy of each quasiparticle is $\Delta$, so the gap in the optical conductivity is $\omega_g = 2\Delta \approx 3.5 T_c$.

 It was once thought that the highest  $T_c$ for a BCS superconductor was around $30^oK$. But in 2001, $MgB_2$ was found to be superconducting at $40^oK$ and is believed to be described by BCS. Some people now speculate that BCS could describe a superconductor with $T_c = 200^oK$.\footnote{Discussion at the KITP program on Quantum Criticality and the AdS/CFT Correspondence, July 2009.}
 
A new class of high $T_c$ superconductors were discovered in 1986 \cite{Bednorz:1986tc}. They are cuprates and the superconductivity is along the $CuO_2$ planes. 
 The highest $T_c$ known today (at atmospheric pressure)  is $T_c = 134^oK$ for a mercury, barium, copper oxide compound.  If you apply pressure, $T_c$ climbs to about 160K.
Another class of superconductors were discovered in 2008 based on iron and not copper \cite{Hosono:2008}. The  highest $T_c$ so far is $56^oK$. These materials are also layered and the superconductivity is again associated with the two dimensional planes.  They are called  iron pnictides since they involve other elements like arsenic  in the nitrogen group of the periodic table.

There is evidence that electron pairs still form in these high $T_c$ materials, but the pairing mechanism is not well understood. Unlike BCS theory, it involves strong coupling. 
Gauge/gravity duality is an new tool to study strongly coupled field theories. In particular, it allows one to  compute dynamical transport properties of strongly coupled systems at nonzero temperature. Condensed matter theorists have very few tools to do this.
We will describe below the first steps toward applying this new tool to better understand high $T_c$ superconductivity. I must stress at the beginning that we are still at the early stages of this endeavor. We will construct simple gravity models and show that they reproduce basic properties of superconductors. But our models are too crude to make detailed comparisons with any real-world material. 

%There is experimental evidence that conformal  field theories play a role in understanding high $T_c$ superconductors. Linear resistivity up to 800K and down to zero if you apply a magnetic field to suppress the superconductivity.

 \section{A gravitational dual}

How do we go about constructing a holographic dual for a superconductor? The minimal ingredients are the following. In the superconductor we need a notion of temperature. On the gravity side, that role is played by a black hole. Recall that in the 1970's Hawking (following work by Bekenstein and others) showed that stationary black holes are thermodynamic objects with a  temperature $T$ related to the surface gravity $\kappa$ via $T=\kappa/2\pi$. In gauge/gravity duality, the Hawking temperature of the black hole is identified with the temperature of the dual field theory\footnote{I will assume that the reader is familiar with the basics of gauge/gravity duality. If not, see other contributions in this book.}. Since gauge/gravity duality traditionally requires that spacetime asymptotically approach anti de Sitter (AdS) space at infinity, we will be studying black holes in AdS. Unlike asymptotically flat black holes, these black holes have the property that at large radius, their temperature increases with their mass, i.e., they have positive specific heat, just like familiar nongravitational systems. There are also planar AdS black holes, which will be of most interest. These black holes always have positive specific heat.

In the superconductor, we also need a condensate. In the bulk, this is described by some field coupled to gravity. A nonzero condensate corresponds to a static nonzero field outside a black hole. This is usually called black hole ``hair".  So to describe a superconductor, we need to find a black hole that has hair at low temperatures, but no hair at high temperatures. More precisely, we need the usual Schwarzschild or Reissner-Nordstrom AdS black hole (which exists for all temperatures) to be unstable to forming  hair at low temperature. At first sight, this is not an easy task.
There are ``no-hair theorems" which say that certain matter fields must be trivial outside a black hole (see, e.g., \cite{Bekenstein:1996pn,Heusler:1996ft}). The idea behind these theorems is simply that matter outside a black hole wants to fall into the horizon  (or radiate out to infinity in the asymptotically flat case). However, there is no general ``no-hair theorem". Each matter field must be considered separately. The result is a set of black hole uniqueness theorems  showing that when gravity is coupled to certain matter fields, stationary black holes are uniquely characterized by their conserved charges: mass, angular momentum and electromagnetic charge. These theorems usually require
 linear matter  fields or scalars with certain potentials $V(\phi)$. Counterexamples to a general no-hair theorem  have been known since the early 1990's. (So our task is not impossible.) For example, it was shown that static Yang-Mills fields can exist outside the horizon \cite{Volkov:1989fi}.

String theory has many ÒdilatonicÓ black holes with scalar hair, and one might be tempted to try to use one of these  to model a superconductor. But if the action includes a term like $e^{2a\phi} F_{\mu\nu}F^{\mu\nu}$, this is doomed to failure.  In this case,
$ F^2$ is a source for $\phi$, so all charged black holes have nonzero $\phi$. This ``secondary hair" is not what we want, since we want the hair to go away at high temperatures. (Theories with more general coupling $f(\phi) F_{\mu\nu}F^{\mu\nu}$ are possible candidates provided $f$ does not have a linear term in $\phi$. However the example we will study uses a standard Maxwell action.) A general argument against AdS black holes developing scalar hair was given by Hertog \cite{Hertog:2006rr}. He considered
 a real scalar field with arbitrary potential $V(\phi)$ (with negative extremum so AdS is a solution), and showed that neutral AdS black holes can have scalar hair if and only if AdS itself is unstable. This is clearly unacceptable.

A surprisingly simple solution to this problem was found by Gubser \cite{Gubser:2008px}. He argued that a charged scalar field around a {\it charged} black hole in AdS would have the desired property. Consider
\be\label{action}
 S=\int d^{4}x \sqrt{-g}\left(R + {6\over L^2} -\frac{1}{4}F_{\mu \nu}F^{\mu \nu} - |\nabla\Psi-i qA\Psi|^2 - m^2|\Psi^2|\right). 
\ee
This is just general relativity with a negative cosmological constant $\Lambda = -3/L^2$, coupled to a Maxwell field and charged scalar with mass $m$ and charge $q$. It is easy to see why black holes in this theory might be unstable to forming scalar hair:
For an electrically charged black hole, the effective mass of $\Psi$ is $m^2_{eff} = m^2 + q^2 g^{tt} A_t^2$. But the last term is negative, so there is a chance that $m^2_{eff}$ becomes sufficiently negative near the horizon to destabilize the scalar field. Furthermore, as one lowers the temperature of a charged black hole, it becomes closer to extremality which means that $g_{tt}$ is closer to developing a double zero at the horizon. This means that $|g^{tt}|$ becomes larger and the potential instability becomes stronger at low temperature.

We will see that black holes in this theory indeed develop scalar hair at low temperature. One might wonder why such a simple type of hair was not noticed earlier. One reason is that this does not work for asymptotically flat black holes. The AdS boundary conditions are crucial. 
 One way to understand the difference is by the following quantum argument. Let $Q_i$ be the initial charge on the black hole. 
 If $qQ_i$ is large enough, even maximally charged black holes with zero Hawking temperature create pairs of charged particles. This is simply due to the fact that the electric field near the horizon is strong enough to pull pairs of oppositely charged particles out of the vacuum via the Schwinger mechanism of ordinary field theory.  The particle with opposite charge to the black hole falls into the horizon, reducing $Q_i$ while the particle with the same sign charge as  the black hole is repelled away. In asymptotically flat spacetime, these particles escape to infinity, so the final result is a standard Reissner-Nordstrom black hole with final charge $Q_f < Q_i$. In AdS, the charged particles cannot escape since the negative cosmological constant acts like a confining box, and they settle outside the horizon. This gas of charged particles is the quantum description of the hair. This quantum process has an entirely classical analog in terms of superradiance of the charged scalar field. 

The four dimensional bulk theory (\ref{action}) is dual to a 2+1 dimensional boundary theory. This is the right context to try to understand the superconductivity associated with two dimensional planes in the high $T_c$ cuprates or iron pnictides. One can also study this theory in five dimensions to describe $3+1$ dimensional superconductors. I should emphasize that at the moment we are not trying to derive the gravitational  theory from string theory. The idea is to find a simple gravity theory with the properties we want, and  analyze it using standard entries in the gauge/gravity duality dictionary. However, we will see later that this simple model can, in fact, be realized   as a consistent truncation of string theory.

Before we proceed, I should comment on the following confusing point. In gauge/gravity duality, gauge symmetries in the bulk correspond to global symmetries in the dual field theory. So although the scalar hair breaks a local $U(1)$ symmetry in the bulk, the dual description consists of a condensate breaking a global $U(1)$ symmetry. Thus strictly speaking, the dual theory is a superfluid rather than a superconductor. The superfluid properties of this model have indeed been investigated \cite{Herzog:2008he,Basu:2008st,Amado:2009ts}. However, one can still view the dual theory as a superconductor in the limit that the $U(1)$ symmetry is ``weakly gauged"\footnote{This means that one imagines that the dual action includes terms like $|\nabla_i - ie A_i)\varphi|^2$ with very small charge $e$.}. In fact, most of condensed matter physics does not include dynamical photons, since their effects are usually small. In particular, BCS theory only includes the electrons and phonons. Electromagnetic fields are usually introduced as external fields, as we will do here. (Unfortunately, our electromagnetic fields will not be dynamical on the boundary.)

\section{Probe Limit}

If one rescales $A_\mu = \tilde     A_\mu/q$  and $\Psi =\tilde      \Psi/q$, then the matter action in (\ref{action}) has a $1/q^2$ in front, so the backreaction of the matter fields on the metric is suppressed when $q$ is large. The limit $q\rightarrow \infty$ with $ \tilde     A_\mu$  and $\tilde      \Psi$ fixed is called the probe limit. It simplifies the problem but retains most of the interesting physics since the nonlinear interactions between the scalar and Maxwell field are retained.   In this section, we will explore this probe limit. 
 We  first discuss the formation of the condensate and then compute the conductivity.
To simplify the presentation, we will drop the tildes.

\subsection{Condensate}

We start with the  planar Schwarzschild anti-de Sitter black hole in four dimensions
\be\label{sads}
ds^2 = - f(r) dt^2 + \frac{dr^2}{f(r)} + r^2 (dx^2 + dy^2) \,,
\ee
where
\be
f = \frac{r^2}{L^2}\left(1 - \frac{r_0^3}{r^3} \right) \,.
\ee
$L$ is the AdS radius, and the Schwarzschild radius $r_0$ determines the Hawking temperature of the black hole:
\be\label{temp}
T = \frac{3 r_0}{4 \pi L^{2}} \,.
\ee
In the probe approximation this metric is a fixed background in which we solve the Maxwell-scalar equations.

We assume a plane symmetric ansatz, 
\be
\Psi = \psi(r), \qquad A_t = \phi(r)
\ee
 With $A_r = A_x = A_y =
0$, the Maxwell equations imply that the phase of $\psi$ must be
constant. Without loss of generality we therefore take $\psi$ to
be real.  The Maxwell-scalar field equations reduce to the following coupled, nonlinear, ordinary differential equations:
\be\label{psieq}
\psi''+\left(\frac{f'}{f}+\frac{2}{r}\right)\psi'+\frac{\phi^2}{f^2}\psi-\frac{m^2}{f}\psi=0
\ee
\be
\phi''+\frac{2}{r}\phi'-\frac{2\psi^2}{f}\phi=0.\label{phieq}
\ee
The key term in the first equation is  $(\phi^2/f^2)\psi$. This comes in with the opposite sign of the mass term and will cause the scalar hair to form at low temperature.

        We first consider the case $m^2 = - 2/L^2$. It might seem strange to make the scalar field tachyonic, but this mass is perfectly allowed in gauge/gravity duality. First note that a tachyonic mass in field theory does not describe particles moving faster than the speed of light. Instead, it usually describes an instability. The value $\psi=0$ is unstable and the field rolls off the potential. However, Breitenlohner and Freedman \cite{Breitenlohner:1982jf} showed that  $AdS_{d+1}$  spacetime is stable even with scalar  fields with $m^2 <0$  provided $m^2 \ge m^2_{BF}$ with
        \be\label{eq:BF}
        m^2_{BF} = - {d^2\over 4L^2}
        \ee
   This is because there is so much volume at large radius that  the positive gradient energy can compensate for a negative $m^2$. Our choice of mass satisfies this bound and in fact  
corresponds to a conformally coupled scalar in $AdS_4$. The standard compactification of 11D supergravity on $S^7$ produces many fields with this mass.

      We now consider the boundary conditions.   At the horizon, one often argues that $\phi = A_t$ must vanish in order for $g^{\mu\nu} A_\mu A_{\nu}$ to remain finite. This is correct, but $A_\mu$ is gauge dependent, so it is not obvious that a diverging vector potential is a problem if the Maxwell field remains finite. A better argument for setting $A_t=0$ at the horizon is the following\footnote{I thank Karl Landsteiner for suggesting this.}. The source for Maxwell's equations in the bulk is, of course, gauge invariant. But  in a gauge in which $ \psi$ is real, the current is just $\psi^2 A_\mu$. Since the current must remain finite at the horizon, we need $A_\mu$ to remain finite, and hence $\phi  = A_t=0$. Even in Einstein-Maxwell theory (without  charged sources) $A_t$ must vanish at a static black hole horizon by the following argument:  For describing thermal properties of the black hole, one should use the Euclidean solution. The Wilson loop of $A_\mu$ around the Euclidean time circle is finite and gauge invariant. If $A_t$ is nonzero at the horizon, the Wilson loop is nonzero around a vanishing circle which implies that the Maxwell field is singular.

There is another constraint that must be satisfied in order for the solution to be smooth at the horizon.
 Mutliplying (\ref{psieq}) by $f$ and evaluating at $r=r_0$ one finds $f' \psi' = m^2 \psi$, so  $\psi(r_0)$ and $\psi' (r_0)$ are not independent. As a result, even though we start with two second order equations which have a four parameter family of solutions, there is only a two parameter subfamily which is regular at the horizon. They can be labelled by 
 $\psi(r_0)$ and $\phi'(r_0)$. (Note that $\phi'(r_0)$ is essentially the electric field at the horizon:  $(\phi')^2 = F_{\mu\nu} F^{\mu\nu}$.)
      
     We now turn to the boundary condition at infinity.  Asymptotically:
      \be\label{asymppsi}
\psi = \frac{\psi^{(1)}}{r} + \frac{\psi^{(2)}}{r^2} + \cdots \,.
\ee
and
\be\label{asympphi}
\phi = \mu - \frac{\rho}{r} + \cdots \,.
\ee
Usually, normalizability requires that the leading coefficient in $\psi$ must vanish. However, since we have chosen a mass close the the BF bound, even the leading term in $\psi$ is normalizable. In this case, one has a choice: One can consider solutions with $\psi^{(1)} =0$ or $ \psi^{(2)} =0$. For definiteness, we will mostly consider standard boundary conditions, $\psi^{(1)} =0$ .

After imposing this asymptotic boundary condition, we have a one parameter family of solutions, which can be found numerically \cite{Hartnoll:2008vx}.  (See \cite{Gregory:2009fj} for an accurate analytic approximation and \cite{Koutsoumbas:2009pa} for an analytic solution of a related model.) We won't present any plots of $\psi(r)$ and $\phi(r)$ since they look rather boring.
 It is easy to see from (\ref{phieq}) that $\phi$ is a monotonic function.  It starts at zero at the horizon and at any local extremum $\phi'' \propto \phi$. So it cannot have a positive maximum or a negative minimum.  If it starts increasing away from the horizon, it continues to increase and asymptotically approaches the constant $\mu$.   $\psi(r)$ does not have to be monotonic. There is a discrete  infinite family of solutions for $\psi(r) $ that satisfy our asymptotic boundary conditions. They can be labelled by the number of times $\psi $ vanishes. It is believed  that only the lowest solution which monotically decreases from $\psi(r_0)$ to zero is stable, although I do not think a stability analysis has been performed yet. Despite the fact that the solutions of interest are simple monotonic functions, they have important consequences for the dual field theory as we now discuss.

     Since we have not started with a consistent truncation of string theory, we do not have a detailed microscopic description of the dual field theory. Nevertheless, basic elements of the gauge/gravity duality dictionary allow us to say the following. The dual theory is a  2+1 dimensional conformal field theory (CFT) at temperature T given by (\ref{temp}). The local gauge symmetry in the bulk corresponds to a  global $U(1)$ symmetry in the CFT.  The asymptotic behavior of the bulk solution determines certain properties of the dual field theory. For example, from (\ref{asympphi}),
$\mu$ is the chemical potential and $\rho$ is the  charge density. It may seem strange that our superconductor has a nonzero charge density. After all, ordinary superconductors are electrically neutral. Perhaps the best analogy is to say that our holographic superconductor is modeling the electrons, but does not include the (positively charged) atomic lattice. Indeed, our model has translational symmetry, and there is no sign of any lattice. From a practical standpoint, we need the charge since neutral black holes in the bulk are not unstable to forming scalar hair. In addition, without the charge (or chemical potential), the dual theory is scale invariant and cannot have a phase transition.

 The dual theory also has an operator charged under the $U(1)$ which is dual to $\psi$.
Since we have chosen $m^2$ close to the BF bound, there are  two possible operators depending on how one quantizes $\psi$ in the bulk\footnote{We will work in the classical limit, but the correspondence applies to the full quantum theory.} \cite{Klebanov:1999tb}. If the modes are defined with the standard boundary condition (faster falloff) for $\psi$ in the bulk,  the dual operator has  dimension two. In this case, a nonzero $\psi^{(1)}$  corresponds to adding a
 source for this operator in the CFT, and a nonzero $\psi^{(2)}$ corresponds to a nonzero expectation value\footnote{This normalization of $\ocal$ differs from that of \cite{Hartnoll:2008vx} by a factor of $\sqrt{2}$.} 
  \be
  O_2 = \psi^{(2)}.
 \ee
   Since we want the condensate to turn on without being sourced,
 we  have set $\psi^{(1)}=0$.  There is an alternative quantization of $\psi$ in the bulk in which the roles of $\psi^{(1)}$ and $\psi^{(2)}$ are reversed. $\psi$ is now dual to  a dimension one operator. If one wishes to study this case, one should impose the boundary condition $\psi^{(2)} =0 $.
  
 We want to know how the condensate $ O_2$ behaves as a function of temperature. Before presenting the results we have to discuss an important scaling symmetry of our problem. In any conformal field theory on $R^n$, one can change the temperature  by a simple rescaling. In the bulk, this is reflected in the statement that the rescaling
 \be\label{rescale2}
 r\rightarrow ar,  \qquad (t,x,y) \rightarrow (t,x,y)/a, \qquad r_0 \rightarrow ar_0
 \ee
 leaves the form of the black hole  (\ref{sads}) invariant with  $f \rightarrow a^2 f$. It is easy to check that the Maxwell-scalar field equations (\ref{psieq},\ref{phieq}) are invariant under this rescaling if $\phi \rightarrow a\phi $ (so $A = \phi dt$ is invariant) and $\psi$ is unchanged $\psi \rightarrow \psi$. Rather than discuss this trivial dependence on temperature which simply reflects the scaling dimension, 
 we are interested in the dependence of a dimensionless measure of the condensate as a function of a dimensionless measure of the temperature. It is convenient to use the chemical potential to fix a  scale and consider $\sqrt { O_2}/\mu$ as a function of $T/\mu$. When one does this, one finds that the condensate is nonzero only when $T/\mu$ is small enough.  Setting $T_c$ equal to the critical temperature when the condensate first turns on,  we get Fig. 1.

   \begin{figure}\begin{center}
\includegraphics[width=.7\textwidth]{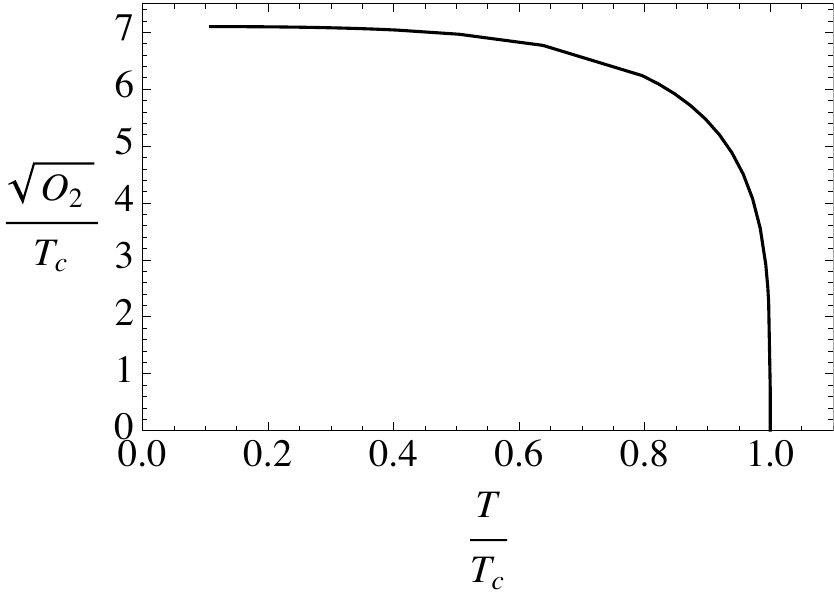}
\caption{The condensate as a function of temperature. The critical temperature is proportional to the chemical potential.}\label{cond2}
\end{center}\end{figure}

This curve is qualitatively similar to that
obtained in BCS theory, and observed in many materials, where the
condensate rises quickly as the system is cooled below the critical temperature and goes to a constant as $T\rightarrow 0$.    
     Near $T_c$, there is a square root behavior $ O_2 = 100T_c^2(1 - T/T_c)^{1/2}$. This is the standard behavior predicted by Landau-Ginzburg theory.  

A nonzero condensate means that the black hole in the bulk has developed scalar hair. 
 One can compute the free energy (euclidean action) of these hairy configurations and compare with the solution $\psi= 0$,  $\phi =  \rho(1/r_0 - 1/r)$ with describes a black hole with the same charge or chemical potential, but no scalar hair. It turns out that the free energy is always lower for the hairy configurations and becomes equal as $T\rightarrow T_c$ \cite{Hartnoll:2008vx}. The difference of free energies scales like $(T_c - T)^2$ near the transition, showing that this is a second order phase transition. Actually, you can compare the free energy at fixed charge or fixed chemical potential. These correspond to two different ensembles. In both cases, the free energy is lower for the hairy configuration.

  We now    generalize to other  masses.
Recall that the asymptotic behavior of a  scalar field of mass $m$ in $AdS_{4}$ is
\be
\psi = {\psi_-\over r^{\lambda_-}} + {\psi_+\over r^{\lambda_+}}+\cdots\label{asymptscalar}
\ee
 where 
 \be\label{lambda}
 \lambda_\pm=\frac{1}{2}\left(3\pm\sqrt{9+4(mL)^2}\right).
 \ee
  For $m^2\ge m^2_{BF}+L^{-2}$, only the mode with $\lambda_+$ falloff is normalizable, and so we interperet $\psi_+=\ocal$, where $\ocal$ is the expectation value of the operator dual to the scalar field, and $\lambda_+$ is the dimension of the operator.  $\psi_-$ is dual to a source for this operator, so we only consider solutions with the standard boundary condition $\psi_- =0$. For $m^2_{BF} \le m^2 < m^2_{BF}+L^{-2}$ one can also consider the alternative boundary condition $\psi_+ =0$, and the dual operator has  dimension  $\lambda_-$.
   
 Fig. 2 shows the results for the condensate for various masses. It is convenient to label the curves in terms of the dimension of the condensate $\lambda$, rather than the mass to distinguish the two possible cases when $m^2 = -2/L^2$. The masses range from the BF bound to $m^2=0$. Similar behavior is found for $m^2 > 0 $ \cite{Kim:2009kb}.
 
\begin{figure}\begin{center}
\includegraphics[width=.7\textwidth]{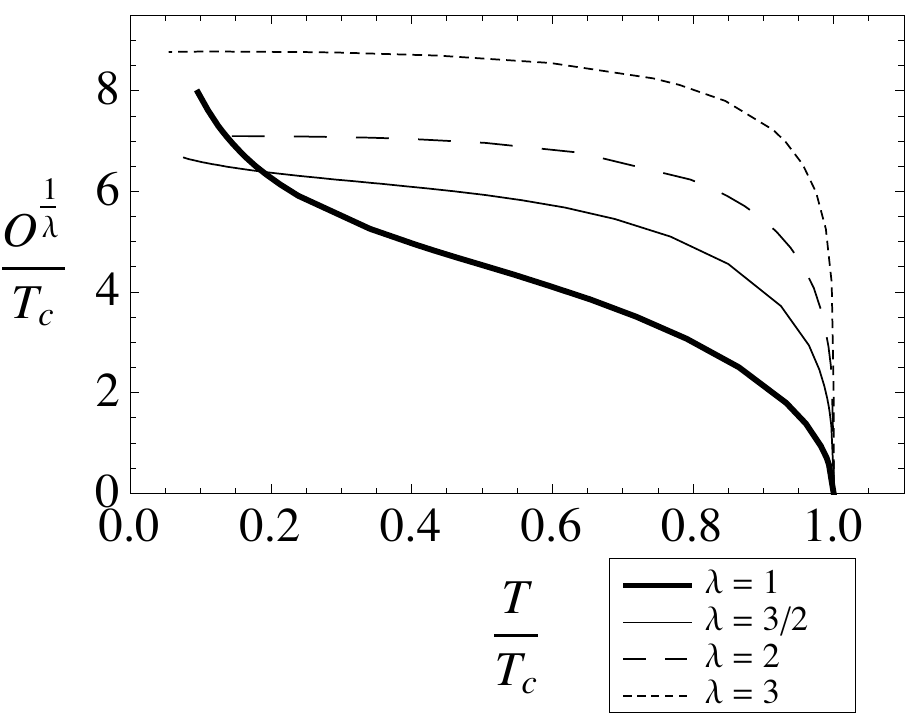}
\caption{Condensates with different dimension, $\lambda$, as a function of temperature.    The condensate tends to increase with $\lambda$. Figure is taken from \cite{Horowitz:2008bn}.}\label{order}
\end{center}\end{figure}

The qualitative behavior is the same as before. In all cases, there is a critical temperature $T_c$ (proportional to the chemical potential) above which the condensate is zero. Near the critical temperature, $O \propto (T_c - T)^{1/2}$.   In all but one case, the condensate saturates as $T\rightarrow 0$. The exceptional case is the dimension one curve which starts to grow at low temperature.
 When the condensate becomes
very large, the backreaction on the bulk metric can no longer be neglected.
We will see later that in the full solution with backreaction, the condensate approaches a finite limit at zero temperature.

It turns out that the ratio $T_c/\mu$ increases as the dimension of the operator decreases. For $\lambda > 3/2$, one can understand this since a smaller $\lambda$ corresponds to a smaller $m^2$ making is easier for the scalar hair to form. However, this continues for $\lambda < 3/2$ when one must increase $m^2$ (and use the alternative boundary conditions) to decrease the dimension. In fact, as observed in  \cite{Denef:2009tp},  $T_c/\mu$ appears to diverge as $\lambda$ approaches the unitarity bound, $1/2$.  The reason for this is not clear. But the lesson is that, at least in this simple model, to have a higher temperature superconductor, one should lower the dimension of the condensate.

\subsection{Conductivity}

We want to compute the optical conductivity, i.e. the conductivity as a function of frequency. By symmetry, it suffices to consider just the conductivity in the $x$ direction. According to the gauge/gravity dictionary, this is obtained by solving for fluctuations in the Maxwell field in the bulk.  Maxwell's equation with zero spatial momentum and time dependence $e^{-i\omega t}$ gives the following equation for $A_x(r)$:

\be
A_x'' +\frac{f'}{f}A_x' + \left(\frac{\omega^2}{f^2}-\frac{2\psi^2}{f}\right)A_x=0 \label{eq:current}
\ee
 We want to solve this with ingoing wave boundary conditions at the horizon, since this corresponds to causal propagation on the boundary, i.e., yields the retarded Green's function \cite{Son:2002sd}. Asymptotically,
\be A_x=A_x^{(0)}+{A_x^{(1)}\over r}+\cdots\ee
The gauge/gravity dictionary says the limit of the electric field in the bulk is the electric field on the boundary: $E_x = - \dot A_x^{(0)}$, and the expectation value of the induced current is the first subleading term: $ J_x = A_x^{(1)}$. From Ohm's law we get:
\be\label{eq:conductivity}
\sigma(\w) = \frac{ J_x }{E_x} = - \frac{  J_x }{\dot A_x^{(0)}} = -\frac{ i  J_x }{\w
A_x^{(0)}} = - \frac{i A_x^{(1)}}{\w A_x^{(0)}} \,.
\ee

The real part of the conductivity is given in Fig. 3 for  the case $\lambda = 2$. Above the critical temperature, the conductivity is constant \cite{Herzog:2007ij}. As you start to lower the temperature below $T_c$ a gap opens up at low frequency. When these curves were first obtained, it was thought that $\Re[\sigma]$  was given by $e^{-\Delta/T}$ at small $\w$, which is what one expects from a BCS type description with an energy gap $\Delta$. This would imply that at $T=0$, $\Re[\sigma]$  should vanish inside the gap. We will see that this is not the case.

\begin{figure}\begin{center}
\includegraphics[width=.7\textwidth]{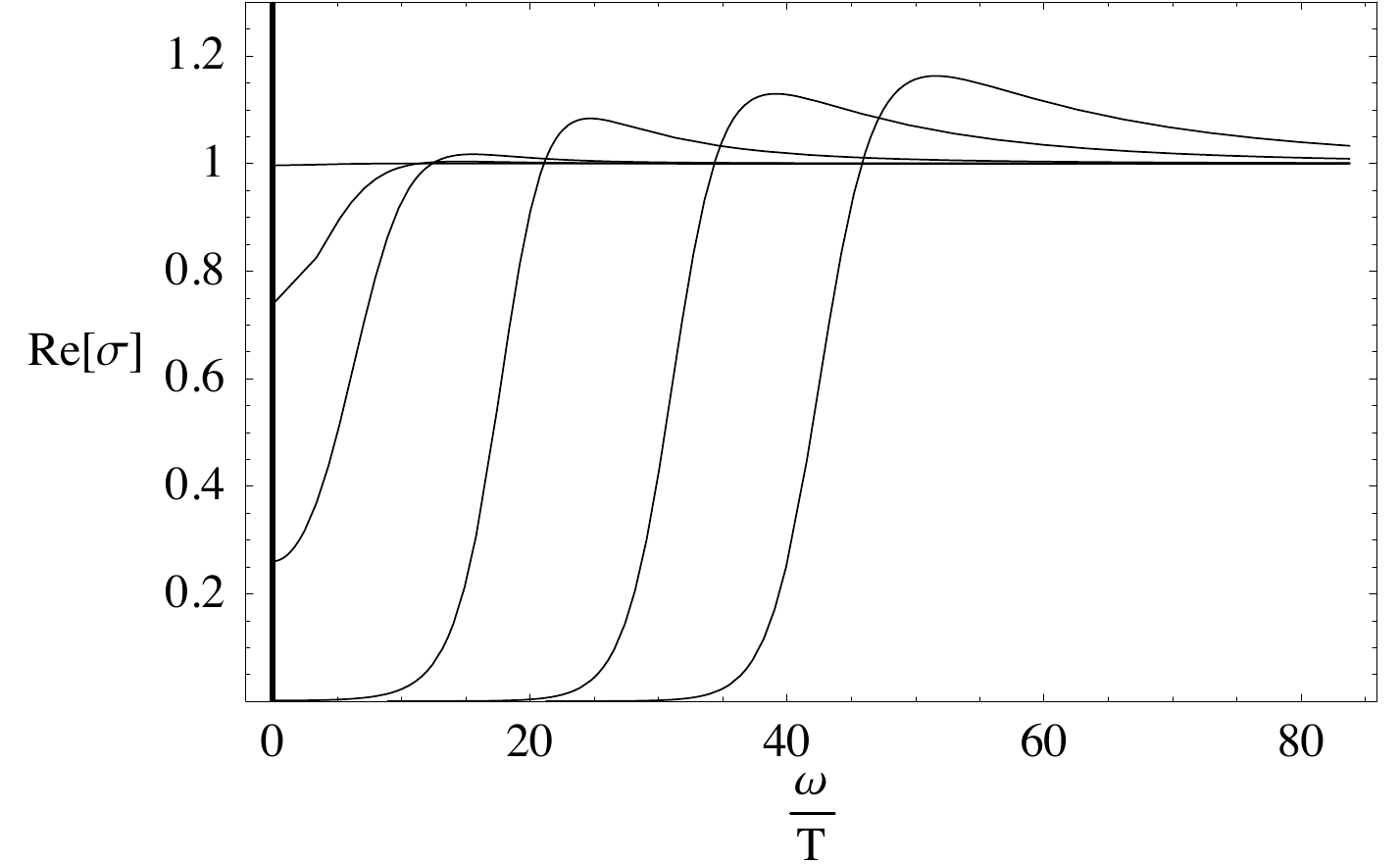}
\caption{The formation of a gap in the real
part of the conductivity as the temperature is lowered below the
critical temperature.  The curves describe successively lower temperatures. There is also a
delta function at $\w = 0$. Figure is for the dimension two condensate and is taken from \cite{Hartnoll:2008vx}.}\label{fig:gap}
\end{center}\end{figure}

There is also a delta function at $\omega = 0$ for all $T < T_c$. This cannot be seen from a numerical solution of the real part, but it can be  
 seen by looking for a pole in $\Im [\sigma]$. A simple argument for this comes from the Drude model of a conductor. Suppose we have charge carriers with mass $m$, charge $e$, and number density $n$ in a normal conductor. They satisfy
 \be
 m {dv\over dt} = eE - m {v\over \tau}
 \ee
 where the last term is a damping term and $\tau$ is the relaxation time due to scattering. The current is $J = env $, so if
 $ E(t) = Ee^{-i\omega t}$,   the conductivity is
 \be
 \sigma(\omega) = {k\tau\over 1 -i \omega \tau} 
 \ee
where $k=ne^2/m$. So 
\be
 \Re[\sigma] = {k\tau\over 1 + \omega^2 \tau^2}, \qquad 
  \Im[\sigma] = {k\omega\tau^2\over 1 + \omega^2 \tau^2}
  \ee
For superconductors, $\tau\rightarrow \infty$, so $\Re[\sigma] \propto \delta(\omega)$ and $\Im[\sigma] \propto 1/\omega$.

A more general derivation comes from the  Kramers-Kronig relations. These 
 relate the real and
imaginary parts of any causal quantity, such as the conductivity,
when expressed in frequency space.  One of the relations is
\be\label{eq:kramers}
{\Im} [\sigma(\w)] = - \frac{1}{\pi} {\mathcal{P}}
\int_{-\infty}^{\infty} \frac{{\Re} [\sigma(\w')] d\w'}{\w'-\w}
\,.
\ee
From this formula we can see that  the real part of the
conductivity contains a delta function,  if and only if the imaginary part has a pole. One finds that there is indeed a
pole in Im$[\sigma]$ at $\omega=0$ for all $T<T_c$.

Fig. 4 shows the low temperature limit of the optical conductivity. The solid line denotes the real part and the dashed line denotes the imaginary part. The pole at $\omega =0$ is clearly visible.

\begin{figure}\begin{center}
\includegraphics[width=.7\textwidth]{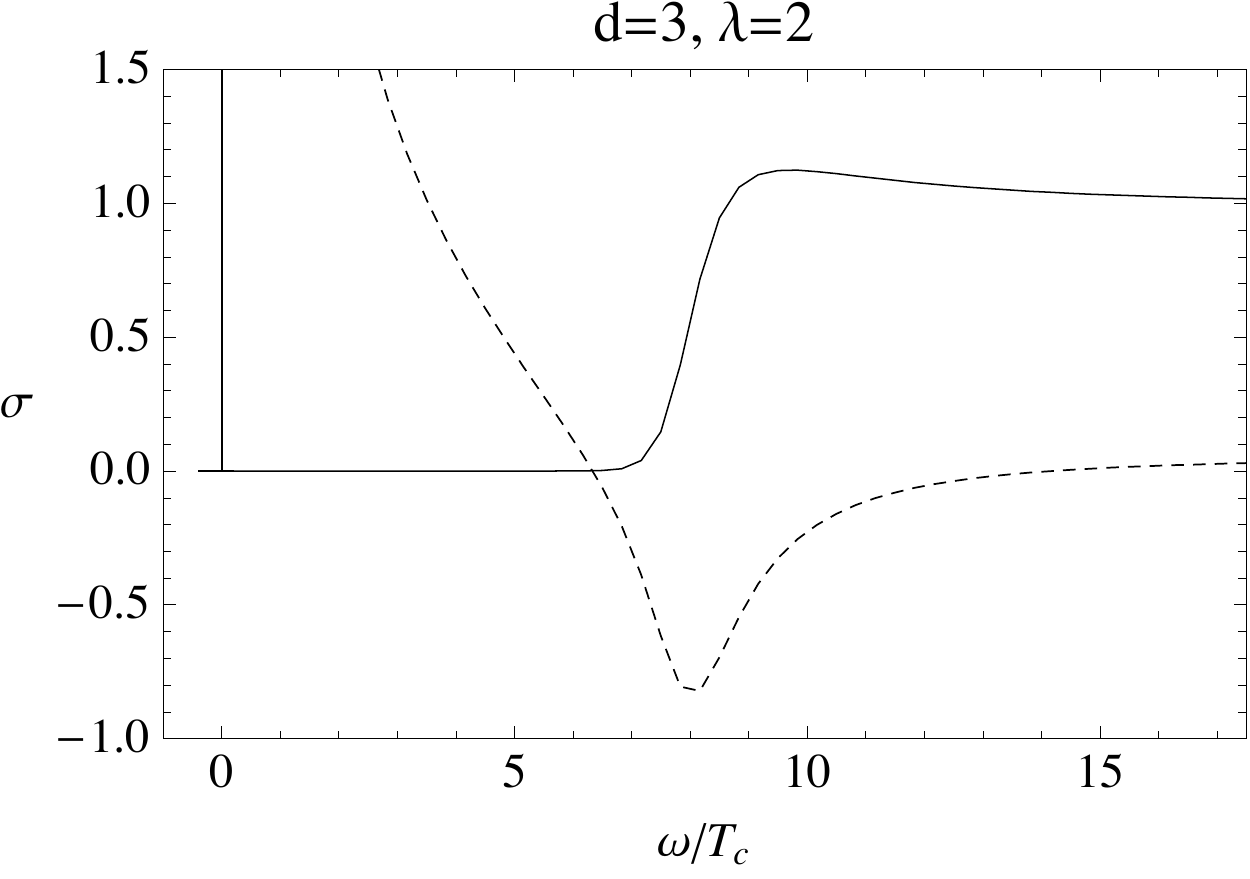}
\caption{The low temperature limit of the optical conductivity for the dimension two condensate. The solid line denotes the real part and the dashed line denotes the imaginary part. Figure is taken from \cite{Horowitz:2008bn}. }\label{fig:gap}
\end{center}\end{figure}

A finite conductivity implies dissipation. In the bulk, this is reflected by the ingoing wave boundary conditions at the horizon. A standard normalizable perturbation of the Maxwell field would decay and get swallowed by the black hole. We maintain a constant amplitude and purely harmonic time dependence by driving the mode with an applied electric field at the boundary. Dissipation normally causes a system to heat up, and indeed if we had included the backreaction of the Maxwell perturbation on the metric, the flow of energy into the horizon would cause the black hole to grow and increase its temperature.

When you approach the BF bound something interesting happens. As shown in Fig. 5, a new spike appears inside the gap. This looks like a bound state of quasiparticles with the binding energy given by distance between the pole and the edge of the gap. However, one must keep in mind that quasiparticles are a weak coupling concept and may not be well defined at strong coupling. We will have more to say about this spike in the next section.

\begin{figure}\begin{center}
\includegraphics[width=.7\textwidth]{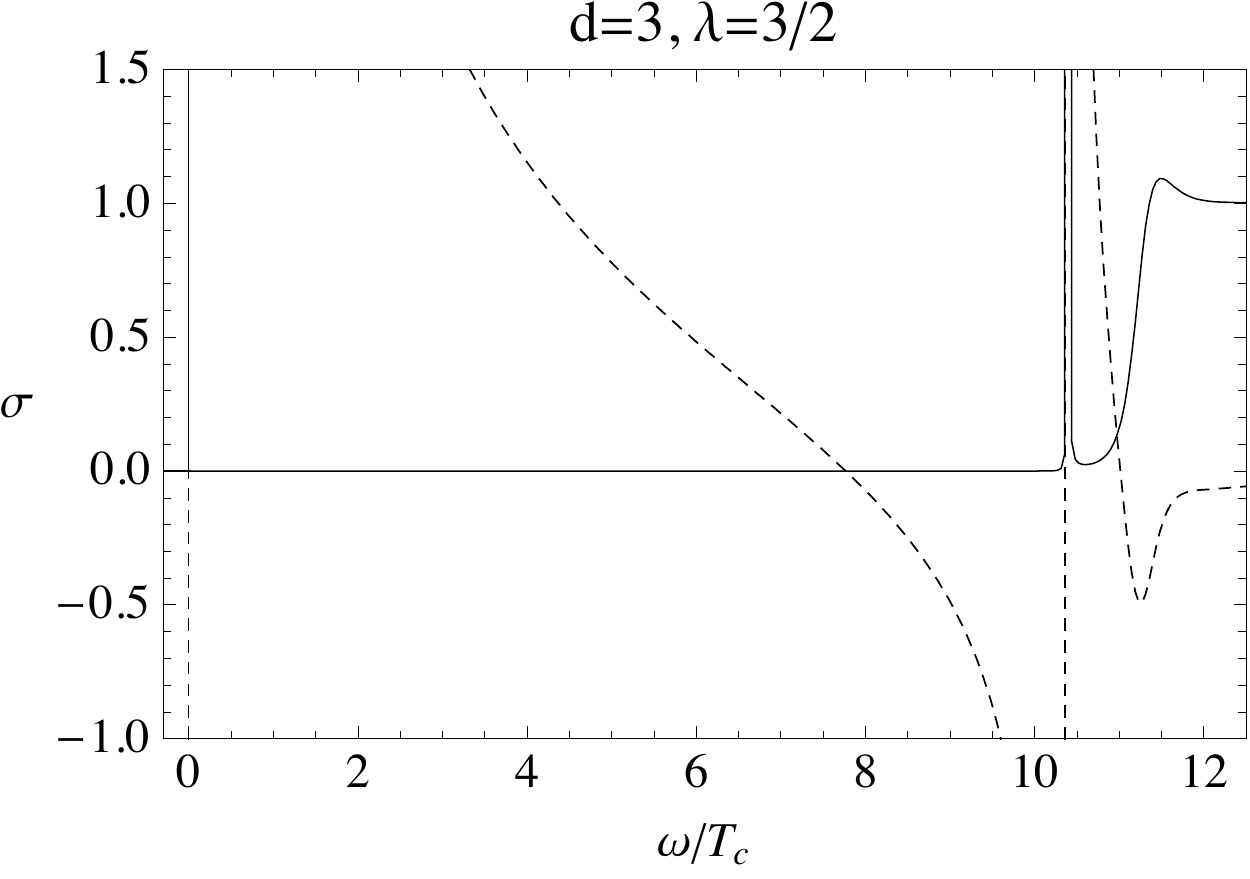}
\caption{The low temperature limit of the conductivity for the dimension 3/2 condensate. Note the extra spike that appears inside the gap. Figure is taken from \cite{Horowitz:2008bn}.}\label{gapspike}
\end{center}\end{figure}

Although we have focussed on the case of $2+1$ dimensional superconductors with a $3+1$ dimensional bulk dual, everything we have done is easily generalized to higher dimensions \cite{Horowitz:2008bn}. The results are qualitatively the same. 

A robust feature that holds in both $2+1$ and $3+1$ superconductors and all $\lambda > \lambda_{BF}$ is that
\be\label{eq:gap}
{\omega_g\over T_c} \approx 8
\ee
with deviations of less than 10\%.\footnote{A few caveats should be noted: If one adds scalar interactions in the bulk by introducing a more general potential $V(\Psi)$, the gap in the low temperature optical conductivity can become much less pronounced, so that this ratio becomes ill defined \cite{Gubser:2008pf}. We will see in the next section that it also becomes ill defined at small $q$. Even when it is well defined, it is modified by higher order curvature corrections in bulk \cite{Gregory:2009fj}.}  This is more than twice the weakly coupled BCS value of $3.5$.  Remarkably, measurements of this ratio in the high $T_c$ superconductors give roughly this value \cite{Gomes:2007}!

%%%%%%%%%%%%%

\section{Full solution with backreaction}

We now discuss the full solution to action (\ref{action}) including the backreaction on the spacetime geometry. 
We start with  the following ansatz for the metric
\be\label{eq:metric}
 ds^2=-g(r) e^{-\chi(r)} dt^2+{dr^2\over g(r)}+r^2(dx^2+dy^2)
\ee
and use the same ansatz for the matter fields as before:
\be
A=\phi(r)~dt, \quad \Psi = \psi(r)
\ee
The equations of motion are\footnote{We will set $L=1$ for the rest of our discussion.}:
\be \psi''+\left(\frac{g'}{g}-\frac{\chi'}{2}+\frac{2}{r} \right)\psi' +\frac{q^2\phi^2e^\chi}{g^2}  \psi  -{m^2\over g}\psi=0      \label{psieom}
\ee

\be\label{phieom}
\phi''+\left(\frac{\chi'}{2}+\frac{2}{r}  \right)\phi'-\frac{2q^2\psi^2}{g}\phi=0
\ee

\be
\chi'+r\psi'^2+\frac{rq^2\phi^2\psi^2e^\chi}{g^2}=0\label{chieom}
\ee

\be\label{geom}
g' + \left(\frac{1}{r}  - { \chi'\over 2}\right) g+\frac{r\phi'^2e^\chi}{4}- 3r+\frac{rm^2 \psi^2}{2}=0
\ee
The first two are the scalar and Maxwell equations as before. The last two are the two independent components of Einstein's equation. (There are three nonzero components of Einstein's equation but only two are independent.) Note that the equations for $g$ and $\chi$ are first order, and $\chi$ is  monotonically decreasing. These equations are invariant under a scaling symmetry analogous to (\ref{rescale2}):
\be\label{rescale}
r \to a r \,, \quad (t,x,y) \to (t,x,y)/a \,, \quad g \to a^2 g \,, \quad \phi \to a \phi
\,.
\ee
When the horizon is at nonzero $r$, this can be used to set $r_0 = 1$.  These equations are also invariant under 
\be\label{rescalet}
e^\chi \to a^2 e^\chi, \quad  t\to at, \quad \phi \to \phi/a
\ee
This symmetry can be used to set $\chi =0$ at the boundary at infinity, so the metric takes the standard AdS form asymptotically.

\begin{figure}
\begin{center}
\includegraphics[width=.7\textwidth]{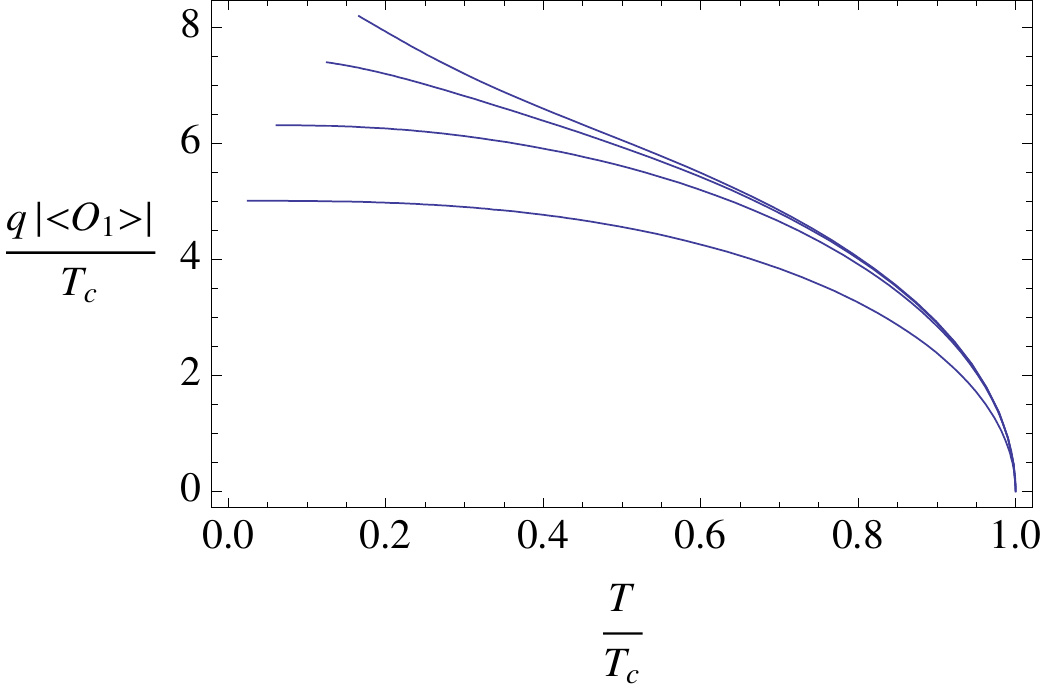}\caption{ From bottom to top, the various curves correspond to $q=1$, 3, 6, and 12. Figure is taken from \cite{Hartnoll:2008kx}.}\label{dimonecond}
\end{center}\end{figure}

 The qualitative behavior of the solutions is similar to the probe limit, i.e., the probe limit indeed captures most of the physics. However there are a few important differences \cite{Hartnoll:2008kx}. 
First,  the apparent divergence in the dimension one condensate at low T is gone. Fig. \ref{dimonecond} shows a plot of $q O_1$ as a function of temperature.  We have multiplied the condensate by the charge $q$ since this is the quantity which is represented by the probe limit at large $q$.  For  $q$ of order one, the curves are similar to the other condensates in the probe limit. But for large $q$, they grow at low temperature. One can show that for all $q$, the condensate still approaches a finite limit at $T=0$.  It is not clear why the dimension one condensate behaves differently in the probe limit from other masses. 
 
The second difference is more surprising. As explained earlier, the origin of the instability responsible for the scalar hair is the coupling of the charged scalar to the charge of the black hole.  It was therefore expected that as $q \rightarrow 0$ the instability would turn off. This is not what happens. A nearly extremal Reissner-Nordstrom AdS black hole remains unstable to forming neutral scalar hair, provided  that $m^2$ is close to the  Breitenlohner-Freedman  (BF) bound.  This means that
there is a new source of instability which can be understood as follows:
An extremal Reissner-Nordstrom AdS black hole has a near horizon geometry $AdS_2 \times R^2$. The BF bound for $AdS_{d+1}$ is $m^2_{BF} = - d^2/4$. So scalars which are slightly above the BF bound for $AdS_4$, can be below  the bound for $AdS_2$. This instability to forming neutral scalar hair is not associated with superconductivity (or superfluidity) since it doesn't break a $U(1)$ symmetry. At most it breaks a $Z_2$ symmetry corresponding to $\psi \rightarrow -\psi$. Its interpretation in the dual field theory is not clear. 

One can make the following general argument for when an extremal Reissner-Nordstrom AdS black hole will be unstable to forming scalar hair \cite{Denef:2009tp}. Consider a scalar field with mass $m$ and charge $q$ in the near horizon geometry of an extremal  Reissner-Nordstrom AdS black hole. Its field equation reduces to a  wave equation in $AdS_2$ with effective mass $m^2_{eff}=(m^2-2q^2)/6$. The $-2q^2$ is the usual coupling to the electric charge and the factor of six comes from the difference between the radius of curvature in $AdS_4$ and $AdS_2$.
The instability to form scalar hair at low temperature is then just the instability of scalar fields below the  BF bound for $AdS_2$: $m^2_{BF} = -1/4$. Thus the condition for instability is 
\be\label{eq:unstable}
 m^2-2q^2<-3/2
 \ee
Of course, the mass must be above the four-dimensional BF bound,  $m^2 > -9/4$, so that the asymptotic $AdS_4$ geometry remains stable. Eq. (\ref{eq:unstable}) is a sufficient condition to guarantee  instability, but it is not necessary.

To compute the conductivity, we again perturb the Maxwell field in the bulk. Assuming zero momentum and harmonic time dependence, the perturbed Maxwell field $A_x$ now couples to the perturbed metric component $g_{tx}$. Physically, this is what one expects from the standpoint of the dual field theory. We are applying an electric field and inducing a current. The current carries momentum, so $T_{tx}$ should be nonzero. This requires that the metric perturbation $g_{tx}$ must also be nonzero\footnote{I thank Hong Liu for this comment.}. However, this also means that one has to solve for the thermal conductivity at the same time as the electrical conductivity. Fortunately, for homogeneous perturbations with harmonic time dependence,  one can solve for $g_{tx}$ in terms of $A_x$ and get
\be
A_x'' + \left[\frac{g'}{g} - \frac{\chi'}{2} \right] A_x'
+ \left[\left(\frac{\w^2}{g^2} - \frac{\phi'^2}{g} \right) e^{\chi} - \frac{2 q^2
\psi^2}{g} \right] A_x  =  0 \,. \label{eq:ax}
\ee
This is similar to (\ref{eq:current}) except for the form of the metric and the extra term $\phi'^2/g$ coming from the metric perturbation $g_{tx}$.
The conductivity is still 
\be\label{oldcond}
\sigma(\omega) = -{i\over \w} {A_x^{(1)}\over A_x^{(0)}}
\ee

 \begin{figure}\begin{center}
 \includegraphics[width=.7\textwidth]{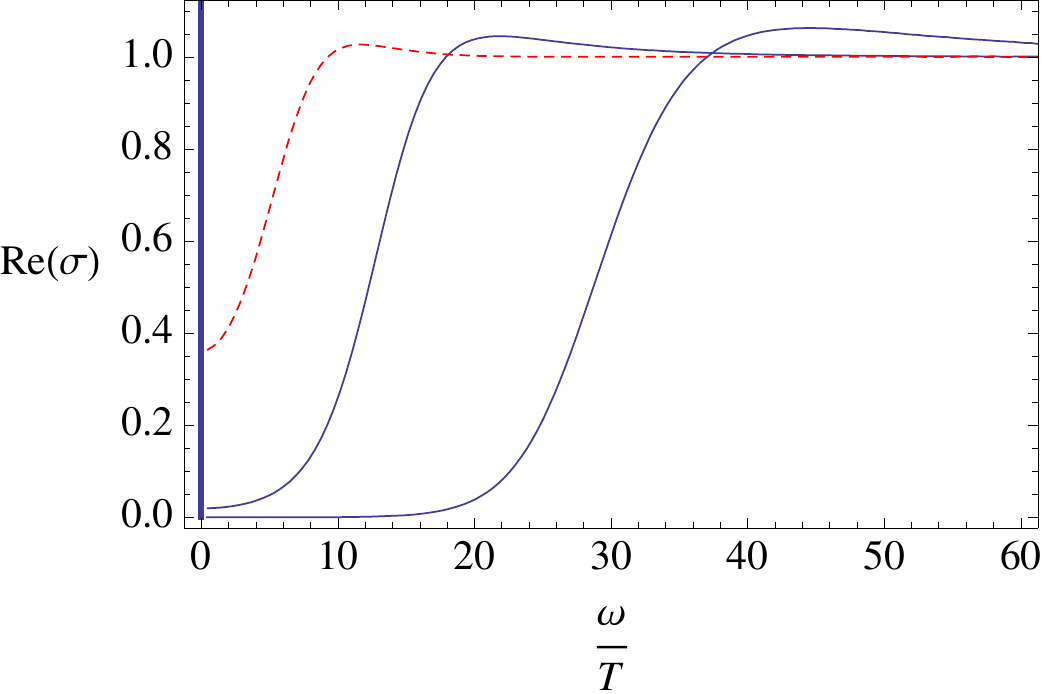}
\caption{Conductivity for the dimension two condensate with $q=3$. The dashed  line is the real part of the conductivity at
$T=T_c$  and the solid lines are the same conductivities at
successively lower temperature $T/T_c = 0.651$, $0.304$. There is a delta function at the origin in all
cases. Figure is taken from \cite{Hartnoll:2008kx}. }\label{fig:gap}
\end{center}\end{figure}

The results for the optical conductivity are qualitatively similar to the probe limit, and shown in Fig. 7. There are a few differences. One is that the conductivity in the normal phase is no longer constant. Another  difference is that in the low temperature limit, the  gap in $\Re \sigma (\omega)$ for small $\omega$ becomes less pronounced at small q. 
The Òrobust featureÓ  $\w_g/T_c \approx 8$  seen in the probe limit continues to hold for $q>3$, but is                 less robust for $q < 3$
mainly because the gap becomes less well defined.
Perhaps the most important difference is that  $\Re \sigma$ now has a delta function contribution at $\omega=0$ even in the normal phase when $T>T_c$. This infinite DC conductivity is not superconductivity, but just a consequence of translational invariance. A translationally invariant, charged system does not have a finite DC conductivity because application of an electric field will cause uniform acceleration. 
One does not see this in the probe limit, since we fixed the gravitational background which  implicitly breaks translation invariance. One indication of this is that we get an electric current without momentum flow as we mentioned earlier. Although there is a delta function at $\omega=0$ for all temperature, its coefficient grows as the temperature is lowered below $T_c$. This indicates the presence of a new contribution coming from the onset of superconductivity.

\subsection{Reformulation of the conductivity}

Equation (\ref{eq:ax}) can be simplified by introducing a new radial variable 
\be
dz = {e^{\chi/2}\over g} dr
\ee
At large $r$, $dz  = dr/r^2$, and we can choose the additive constant so that $ z = -1/r + O(1/r^2)$.
Since $g$ vanishes at least linearly at a horizon and $\chi$ is monotonically decreasing, the horizon corresponds to $z=-\infty$. In terms of $z$, (\ref{eq:ax}) takes the form of a standard Schr\"odinger equation with energy $\omega^2$:
\be\label{schr}
-A_{x,zz} + V(z) A_x = \omega^2 A_x
\ee
where
\be\label{potential} 
V(z) = g[\phi_{,r}^2  + 2q^2 \psi^2  e^{-\chi}]
\ee
From the known asymptotic behavior of the solution near infinity, one can show that $V(0) = 0$ if the dimension of the condensate, $\lambda$, is greater than one, $V(0)$ is a nonzero constant if $\lambda =1$, and $V(z)$ diverges as $z\rightarrow 0$ if $1/2 < \lambda < 1$. One can also show  that the potential always vanishes at the horizon \cite{Horowitz:2009ij}.

We want to solve (\ref{schr})  with boundary conditions at $z = -\infty$ corresponding to waves propagating to the left. This corresponds to ingoing boundary conditions at the horizon which is needed to extract causal results. The easiest way to do this is to first extend the definition of the potential to all $z$ by setting $V=0$ for $z>0$. An incoming wave from the right  will be partly transmitted and partly reflected by the potential barrier. Since the transmitted wave is purely ingoing at the horizon, this satisfies our desired boundary conditions.  Writing the solution for $z>0$ as $A_x = e^{-i\omega z} + \R e^{i\omega z}$, we clearly have $A_x(0) = 1+\R$ and $A_{x,z}(0) = - i\omega(1-\R)$. In terms of $z$, $ A_x^{(1)} = - A_{x,z} (0)$, so from (\ref{oldcond})
\be\label{cond}
\sigma(\omega) =  {1-\R\over 1+\R}
\ee
{\it The  conductivity is directly related to the reflection coefficient, with the frequency simply giving the incident energy \cite{Horowitz:2009ij}!} The qualitative behavior of $\sigma(\omega)$ is now clear. Let us first assume that the dimension of the condensate is $\lambda\ge 1$ so that $V$ is bounded. At frequencies below the height of the barrier, the probability of transmission will be  small, $\R$ will be  close to one, and $\sigma(\omega)$ will be  small. At frequencies above the height of the barrier, $\R$ will be very small and $\sigma(\omega) \sim 1$ (the normal state value). Clearly the size of the gap in $\sigma(\omega)$ is set by the height of the barrier: $\w_g \sim \sqrt {V_{max}}$. 
The case $1/2< \lambda < 1$ is qualitatively similar. Even though the potential is not bounded, $\sqrt{V}$ is integrable, so there is still tunneling through the barrier. 

\begin{figure}
\begin{center}\label{q10potential}
\includegraphics[width=.7\textwidth]{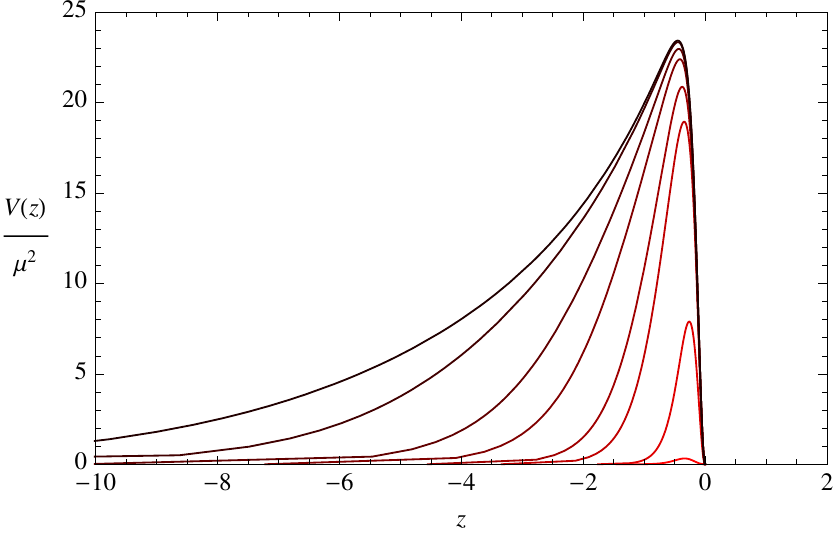}\caption{Schr\"odinger potential for $\lambda=2, q=10$. The top curve as $T=0$ and the bottom curve has $T = T_c$.  Figure is taken from \cite{Horowitz:2009ij}.}\label{pot}
\end{center}\end{figure}

 As one lowers the temperature, the condensate $\psi$ increases and the potential becomes both higher and wider (see Fig.  \ref{pot}).  This causes an increasing exponential suppression which was erroneously interpreted as evidence for Re $\sigma\sim e^{-\Delta/T}$ in the original papers on holographic superconductors. However,  as $T\rightarrow 0$, the potential approaches a finite, limiting form which still vanishes at $z = -\infty$. This means that there will always be nonzero tunneling probability and hence a nonzero conductivity even at small frequency. In other words, there is no hard gap in the optical conductivity at $T=0$. This conclusion continues to hold if the $m^2 |\Psi|^2$ term in our bulk lagrangian is replaced by a general potential $V(\Psi)$. The fact that there is not a hard gap is consistent with calculations of the specific heat which show that it vanishes as a power law at low temperature and not exponentially.

As discussed in section 3.2,  in order for $\Re \sigma$ to have a delta function at $\omega = 0$ 
representing the infinite DC conductivity, one needs  ${\rm Im}\, \sigma$ to have a pole at $\omega =0$. It is easy to see that this is indeed the case, for any positive 
Schr\"odinger  potential $V(z)$ that vanishes at $z=-\infty$.  Imagine solving (\ref{schr}) with $\omega = 0$, and $A_x = 1$ at $z=-\infty$. (This represents the 
normalizable solution and is the zero frequency limit of the ingoing wave boundary condition.) Since $A_{x,zz} > 0$, the solution will be monotonically increasing. At $z=0$, $A_x^{(0)} = A_x(0)$ and $A_x^{(1)} = - A_{x,z}(0)$. These are both real and nonzero. From 
(\ref{oldcond}) it then follows that ${\rm Im}\, \sigma$ has a pole at $\omega =0$. Even in the normal phase when $\psi=0$, the potential is nonzero due to the contribution from the electric field. This is precisely the term which arises due to the backreaction on the metric and is absent in the probe approximation. This is another way to see why the probe limit does not have a delta function at $\omega = 0$ in the normal phase while the full backreacted solution does. Note that the Schr\"odinger potential remains finite in the probe limit $q\rightarrow \infty$, since $q\psi$ is held fixed.

This approach also explains the spike in the conductivity that was seen in Fig. 5.  At low frequency, the incoming wave from the right is almost entirely reflected. If the potential is high enough, one can raise the  frequency so that about one wavelength fits between the potential and $z=0$. In this case, the reflected wave can interfere destructively with the incident wave and cause its amplitude at $z=0$ to be exponentially small. This produces a spike in the conductivity. If one can raise the frequency so that two wavelengths fit between the potential and $z=0$ one gets a second spike, etc. More precisely, using standard WKB matching formula,
spikes will occur when there exists  $\omega$ satisfying
\be
\int_{-z_0}^0 \sqrt{\omega^2 - V(z)} dz + {\pi\over 4} = n\pi
\ee
for some integer $n$, where $V(-z_0) = \omega^2$.  It is now clear that the spike which we saw in the probe approximation with $m^2$ saturating the BF bound, also appears in the full backreacted solution, and for some $m^2$ slightly above the BF bound.
 When the spikes were first seen, it was speculated that they corresponded to vector normal modes of the hairy black hole. It is now clear that they are not true normal modes even at $T=0$, since $A_x$ does not actually vanish at infinity.  The actual modes all have complex frequency and correspond to familiar quasinormal modes. In other words, there are no bound states in this potential (with boundary condition $A_x=0$ at $z=0$) since the potential vanishes at $z=-\infty$.

\section{Zero temperature limit}

The extremal Reissner-Nordstrom AdS black hole has large entropy at $T =0$. If this was dual to a condensed matter system, it would mean the ground state was highly degenerate.
 The extremal limit of the hairy black holes dual to the superconductor is not like 
 Reissner-Nordstrom. It has zero horizon area, consistent with a nondegenerate ground state \cite{Horowitz:2009ij}. One might wonder how this could be the limit of the $T>0$ black holes since we have seen that for all $T\ne 0$, one can scale $r$ so that $r_0 = 1$. The point is simply that the dimensionless ratio  $r_0/\mu \rightarrow $     0 as $ T \rightarrow      0$.  
 
 The extremal limit also has zero charge inside the horizon (except in the case $q=0$ when the scalar hair cannot carry the charge). This is an expected consequence of the horizon having zero area. If the black hole tried to keep a nonzero charge as the temperature was lowered and its horizon shrunk to zero, the electric field on the surface would grow arbitrarily large. Eventually this would pair create charged particles via the Schwinger mechanism and neutralize the black hole. This is a quantum argument involving pair creation, but it has a classical analog in terms of a superradiant instability for charged scalar fields.

The near horizon behavior of the zero temperature solution depends on the mass and charge of the bulk scalar field. We will discuss two classes of solutions below. In both cases, one can solve for the solutions analytically near $r=0$. These leading order solutions depend on a free parameter which can be adjusted so that the solution asymptotically satisfies the desired boundary condition.
 In the cases we study, the solution is typically not smooth at $r = 0$, but since they are the limit of smooth black holes, they are physically allowed. If one modifies the potential for $\Psi$ so that it has more than one extremum, then smooth, zero temperature solutions exist in which $\Psi $ rolls from one extremum to another \cite{Gubser:2008pf}.

\subsection{$m^2 = 0$}

This corresponds to a  marginal, dimension three operator  developing a nonzero expectation value in the dual superconductor. To determine the leading order behavior near $r=0$,
 we try an ansatz
\be
\phi = r^{2+\a},\quad \psi = \psi_0 - \psi_1 r^{2(1+\a)}, \quad \chi =\chi_0 - \chi_1 r^{2(1+\a)}, \quad
g=r^2(1 - g_1r^{2(1+\a)})
\ee
We have used the scaling symmetries (\ref{rescale}) and (\ref{rescalet}) to set the coefficients in $\phi$ and $g$ to one. Substituting this into the field equations  and equating the dominant terms for small $r$ (assuming $\a > -1$),
one finds:
\be
q\psi_0 = \left({\a^2 + 5\a + 6\over 2}\right)^{1/2}, \quad \chi_1 = {\a^2 + 5\a + 6\over 4(\a + 1)}e^{\chi_o}
\ee
\be
g_1 =  {\a + 2\over 4} e^{\chi_o}, \quad \psi_1 = {q e^{\chi_o} \over 2(2\a^2 + 7\a +5)}\left({\a^2 + 5\a + 6\over 2}\right)^{1/2}
\ee

 One can now numerically integrate this solution to large radius and adjust $\a$ so that the solution for $\psi $ is normalizable. One finds that this is possible provided $ q^2 > 3/4$. This is consistent with the  condition for instability (\ref{eq:unstable}). The value of $\a$ depends weakly on $q$ (see Fig. 9). In all cases, $| \a| < .3$.  Fig. 10 shows the results for $\psi$ and $g$ for the zero temperature solution, and shows how the $T>0$ solutions approach it as $T\to 0$.

\begin{figure}
\begin{center}
\includegraphics[width=.7\textwidth]{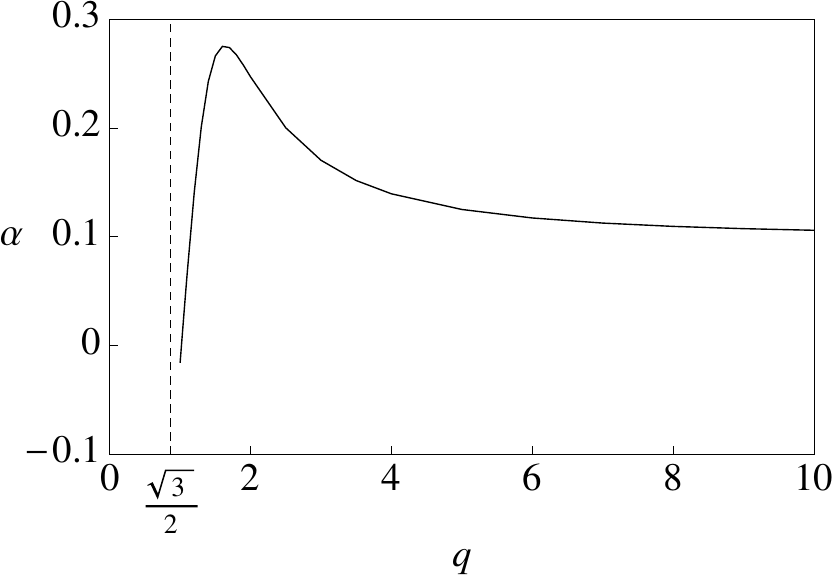}\caption{Values of $\alpha$ for various charges. Note that it approaches a constant as $q\rightarrow \infty$. Figure taken from \cite{Horowitz:2009ij}.}
\end{center}\end{figure}

\begin{figure}
\begin{center}
\includegraphics[width=.45\textwidth]{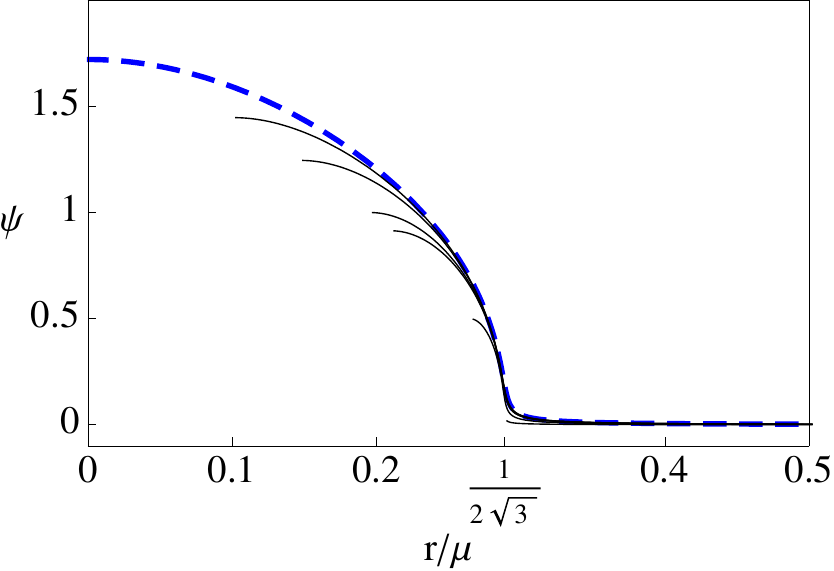}\hspace{0.2cm}
\includegraphics[width=.45\textwidth]{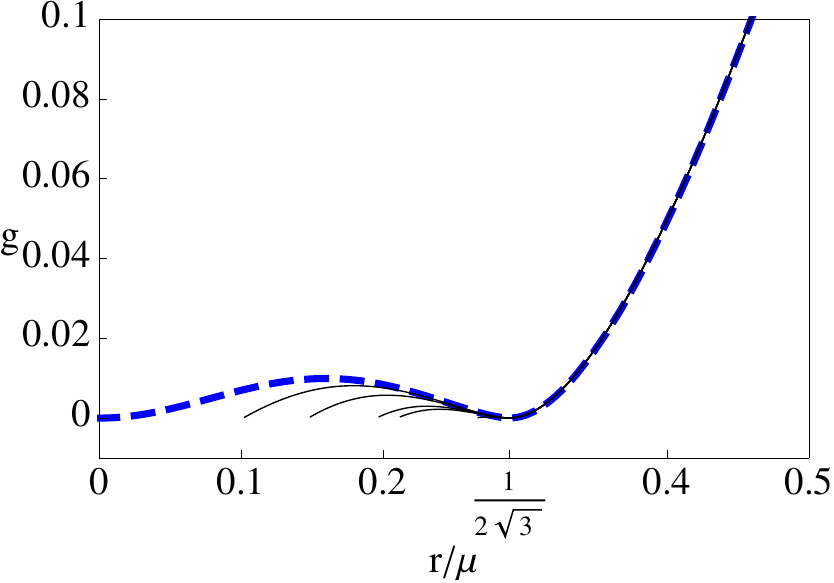}\caption{Zero temperature, $\lambda = 3$ and $q=1$ solution (dashed line), compared to successively lower temperature hairy black holes (solid lines.) Note that $g$ almost has a double zero at $r/\mu=1/2\sqrt{3}$ which is the extremal horizon for  Reissner-Nordstrom AdS. In the limit $q\rightarrow \sqrt{3}/2$ the solution becomes extremal Reissner-Nordstrom AdS with all hair behind the horizon. Figure is taken from \cite{Horowitz:2009ij}.}\label{d3cond}
\end{center}\end{figure}
 
 Near $r=0$, $\chi$ approaches a constant and $g=r^2$. Thus the metric approaches $AdS_4$ with the same value of the cosmological constant as the asymptotic region. The extremal horizon is just the Poincare horizon of $AdS_4$. The scalar field approaches a constant and the Maxwell field vanishes. In terms of the dual field theory, this means that the full conformal symmetry is restored in the infrared. 
 
These solutions are not singular at $r=0$ since all curvature invariants remain finite. 
They can be viewed as  static, charged domain walls. 
Even though they are not singular, when $\a \ne 0$ the solutions are not $C^\infty$ across $r=0$. Some derivatives of the curvature will blow up. However, there is a
 special value of the charge, $q=1.018$,  where $\a =0$. This solution is completely smooth across the  horizon. It is not clear what the significance of this value is for the dual field theory. 

\subsection{$ q^2 > |m^2|/6 \ \ \ (m^2 < 0)$}

%We will call this the large charge case (even though it includes some $q<1$).
We try the following  ansatz near $r=0$:  

\be\label{qansatz}
 \psi=A (-\log r)^{1/2},\quad ~g=g_0 r^2 (-\log  r),\quad ~\phi=\phi_0r^\beta (-\log  r)^{1/2}
\ee
where we have used the radial scaling symmetry (\ref{rescale}) to set an arbitrary length scale in the logarithm to one. 
The behavior of $\chi$ is determined by (\ref{chieom}) and whether we expect $r\psi'^2$ or $rq^2\phi^2\psi^2e^\chi/g^2$ to dominate. 
 We assume $r\psi'^2$ dominates, so
\be e^\chi= K(-\log r)^{A^2/4} \label{large_q_chi}
\ee
with $K$ a constant of integration. The second term in (\ref{chieom}) is indeed negligible provided $\beta >1$.
Equating the dominant terms in the equations of motion leads  to:
\be\label{largeqconst}
 A=2, \quad g_0=-\frac{2}{3}m^2, \quad \beta=-\frac{1}{2}\pm\frac{1}{2}\left( 1-\frac{48 q^2}{m^2}\right)^{1/2}
 \ee

It is clear that for appropriate metric signature we need positive $g_0$ which tells us this ansatz is only appropriate for negative $m^2$. Since we require $\beta>1$, only the plus sign in (\ref{largeqconst}) is allowed and we require
 \be q^2>-m^2/6\ee 
  With these restrictions, our near horizon solution is
 \be\label{qansatzz}
 \psi=2(-\log r)^{1/2},\quad ~g= (2m^2/3) r^2 \log  r,\quad   e^\chi=-K\log r
\ee
\be
~\phi=\phi_0r^\beta (-\log  r)^{1/2},
\ee
The one remaining  free parameter is $\phi_0$. This parameter can be tuned so that  the asymptotic boundary condition on $\psi$ is satisfied.

The horizon at $r=0$ has a mild singularity. The scalar field diverges logarithmically and the metric takes the form (after rescaling $t$)
\be\label{nhmetric}
ds^2 = r^2(-dt^2  + dx_idx^i) + {dr^2\over g_0 r^2(-\log r)}
\ee
Notice that Poincare invariance is restored near the horizon, but not the full conformal invariance. (Some early indications of emergent Poincare symmetry were found in \cite{Gubser:2008pf}). 
Introducing a new radial coordinate, $\tilde r= -2(-\log r)^{1/2}/g_0^{1/2}$ the metric becomes
\be
ds^2 = e^{-g_0\tilde r^2/2}(-dt^2  + dx_idx^i) + d\tilde r^2
\ee
near the horizon which is now at $\tilde r = -\infty$.

\section{Adding magnetic fields}

As we mentioned in the introduction, one of the characteristic properties of superconductors is that they expel magnetic fields. However, a
 superconductor can expel a magnetic field only up to a point. A sufficiently strong  field will destroy the superconductivity. One defines the thermodynamic critical field $B_c$ by setting the energy it takes to expel the magnetic field equal to the difference in free energy between the normal and superconducting states:
\be
{B_c^2 (T)\ V \over 8\pi} = F_n(T) - F_s(T)
\ee
where $V$ is the volume. Superconductors are divided into two classes depending on  how they make the transition from a superconducting to a normal state as the magnetic field is increased. In type I superconductors, there is a first order phase transition at $B=B_c$, above which magnetic field lines penetrate uniformly and the material no longer superconducts.  In  type II superconductors, there is a more gradual second order phase transition.  The magnetic field starts to penetrate the superconductor in the form of vortices with quantized flux when $B=B_{c1} < B_c$.   The vortices become more dense as the magnetic field is increased, and at an upper critical field strength, $B=B_{c2} > B_c$, the material ceases to superconduct.

We now show that our 2+1 dimensional holographic superconductors are type II. (Interestingly enough, the high $T_c$ superconductors are also type II.) We will view this superconductor as a thin film superconductor in 3+1 dimensions with a perpendicular magnetic field. 
Suppose  the 2+1 dimensional sample is a disk of radius $R$.
In order for the disk to expel the magnetic field, it must produce a current circulating around the perimeter.  Solving Maxwell's equations, this current will expel a field not only in the area $\pi R^2$ of the disk but in a larger volume of size $V_3 \sim R^3$.  Since the difference in free energies between the normal and superconducting state scales like $R^2$, in the large $R$ (thermodynamic) limit, the superconductor does not have enough free energy available to expel a magnetic field. Thus magnetic fields of any non-vanishing strength will penetrate a thin superconducting film and $B_c=0$.

To show that the holographic superconductor is type II, we start in the normal phase with a large applied magnetic field, and slowly lower $B$. We will see that parts of  the system start to become superconducting at a value $B_{c2} > 0$ \cite{Albash:2008eh,Hartnoll:2008kx}. To model the normal phase with an applied magnetic field, we use a black hole in AdS with both electric and magnetic charges (but no scalar hair). The magnetic field in the bulk corresponds to applying a background magnetic field on the boundary. The metric takes the form (\ref{eq:metric}) with $\chi=0$ and
\be\label{eq:dyonic}
g(r) = r^2 - \frac{1}{4r r_0} \left(4 r_0^4 + \rho^2 +B^2\right) + \frac{1}{4r^2}(\rho^2 + B^2)  \ .
\ee
The black hole temperature is
\be\label{eq:dyonictemp}
T = \frac{12 r_0^4 - \rho^2 - B^2}{16 \pi r_0^3 } \ .
\ee
The vector potential is 
\be
A = \rho \left( \frac{1}{r_0}- \frac{1}{r} \right)\, dt +  Bxdy .
\ee

Since we are interested in the onset of superconductivity which is a second order phase transition, the condensate will be small and we can ignore the backreaction of $\psi$ on the metric. (Note that this is different from the probe approximation since the backreaction of the Maxwell field is included here.) Since the gauge potential now depends on $x$, we separate variables
as follows
\be
\psi(r,x,y) = R(r) X(x)e^{iky}
\ee
The equation for $X(x)$ turns out to be a simple harmonic oscillator centered at $k/qB$ with width $(qB)^{-1/2}$. The eigenvalues are $2qB(n + {1\over 2})$ but only the lowest mode is expected to be stable. Note that the momentum $k$ shifts the origin of the harmonic oscillator, but does not change its energy. This is just the degeneracy of Landau levels. The radial equation becomes:
\be
(r^2gR')' + \Bigl[{q^2\rho^2\over g}\Bigl({r\over r_0} -1 \Bigr) + qB -m^2 r^2 \Bigr]R = 0
\ee
One can now choose $B$  so that there is a normalizable solution. This defines the upper critical field $B_{c2}$. One finds that  $B_{c2}$ is a decreasing function of temperature and vanishes at $T = T_c$. This is what one expects: At lower temperatures, a superconductor can support a larger magnetic field.  Due to the degeneracy in $k$, the actual condensate involves a superposition of these modes for all $k$ and forms a lattice of vortices.
 Maeda et al. \cite{Maeda:2009vf} have recently studied the free energy as a function of the two parameters which govern the shape of the vortex array and shown that the minimum of the free energy at long wavelength corresponds to a triangular array  which is what is predicted by the  Landau-Ginzburg model.   

\subsection{London equation}

As mentioned in section 2, the boundary theory does not have dynamical Maxwell fields, so one cannot actually see the Meisner effect. However, one can show that in the superconducting state, static magnetic fields induce currents whose backreaction would cancel the applied magnetic field \cite{Maeda:2008ir,Hartnoll:2008kx}.  In fact, one can show that these currents are proportional to $A_i$ and thus reproduce London's equation.

To make life easier, in this section we shall work in the probe limit ($q \to \infty$) in which the metric is kept fixed to be simply the Schwarzschild AdS black hole (\ref{sads}).   Assume that we have solved for $A_t$ and $\psi$ in this background as described in section (3.1).  We then wish to add a static perturbation $A_x$ that has nonzero momentum. We will take the momentum to lie in a direction orthogonal
to $A_\mu$, so we can consistently perturb the gauge field without sourcing any other fields. This perturbation corresponds to adding a static magnetic field in the bulk, and the asymptotic value of this magnetic field corresponds to the magnetic field added to the boundary theory. As before, the first subleading term in $A_x(r)$ corresponds to the induced current. 

Rather than doing this calculation explicitly, we can obtain the answer by relating it to the calculation in section 3.2. Let $A_x$ have both
 momentum and frequency dependence of the form $A_x \sim e^{-i \omega t + i k y}$. 
The radial  equation for $A_x(r)$ reduces to
\be
 (f A_x')' + \left( \frac{\omega^2}{f} - \frac{k^2}{r^2} \right) A_x = 2  \psi^2 A_x \ ,
\label{probemaxwell}
\ee
where $'$ denotes differentiation with respect to $r$.  We studied this equation with $k=0$ in section 3.2, and found a pole in $\Im \sigma[\omega]$ at $\omega = 0$. Let the residue of this pole be $n_s$, the superfluid density. From  (\ref{eq:conductivity}) this means that $A_x^{(1)} = -n_s A_x^{(0)}$. But $J_x = A_x^{(1)}$ and $A_x^{(0)}$ is just the vector potential in the boundary theory, so
\be
J_x (\omega, k=0) = -n_s A_x(\omega, k=0)
\ee
 It is clear from the form of eq. (\ref{probemaxwell}) that the limits $\omega \to 0$ and $k \to 0$ must commute.  
To compute $n_s$, we simply set both $\omega$ and $k$ to zero and solve (\ref{probemaxwell}). 
Thus for small $k$, we directly obtain the  London equation 
\be
  J_i(\omega, k)  = - n_s A_i(\omega, k)
\ee
where we have used the isotropy of space. This equation is clearly not gauge invariant. It is expected to hold in a gauge where $\nabla_i A^i = 0$.  
%It is supposed to be valid when $\omega$ and $k$ are small compared to the scale at which the system loses its superconductivity.  In our case, that scale is $\langle  \ocal_i \rangle$. 

\subsection{Correlation length}

The gauge/gravity dictionary says that the retarded Green's function (for $J_x$) in the dual field theory is  
\be
G^R(\omega,k) = {A_x^{(1)}(\omega,k)\over A_x^{(0)}(\omega,k)}
\ee
We can define a correlation length $\xi$ by expanding
\be
\Re \ G^R(0,k) = -n_s(1 + \xi^2 k^2 + \cdots)
\ee
By solving (\ref{probemaxwell}) numerically, including the $k$ dependence, one can compute this correlation length. One finds that it diverges near the critical temperature like   $\xi T_c \approx 0.1(1-T/T_c)^{-1/2}$ \cite{Horowitz:2008bn}. A similar divergence is found in a correlation length obtained from fluctuations in the condensate \cite{Maeda:2008ir}.

\subsection{Vortices}

As we discussed above, at the onset of superconductivity when $B=B_{c2}$, there is a lattice of vortices.  However at this point, the condensate is  small everywhere and the equations could be linearized. It is of interest to find the bulk solution describing a vortex lattice away from $B=B_{c2}$. The first step is to find the single vortex solution. This corresponds to a something like a cosmic string stretching from the black hole horizon to infinity. However there is a crucial difference between a cosmic string and the dual of a vortex. A standard cosmic string  has a fixed proper radius. This means that its size in the $x,y$ coordinates  used in (\ref{eq:metric}) goes to zero at large $r$. In terms of the boundary theory, this is a point-like object. We want a solution in which the cosmic string is a fixed size in the $x,y$ coordinates, so it has a finite radius in the superconductor. This looks more like a funnel in the bulk with a proper radius that grows  as one increases $r$. 

Such a solution has recently been found in the probe limit \cite{Montull:2009fe}.\footnote{See  \cite{Albash:2009iq} for another approach.} Let us write the background metric using polar coordinates for the flat transverse space:
\begin{equation}
ds^2 = - f(r) dt^2 + \frac{dr^2}{f(r)} + r^2 \left(d\tilde r^2 + \tilde r^2 d\varphi^2
\right) \,.
\end{equation}
Assume an ansatz
\be
\Psi = \psi(r,\tilde r)e^{in\varphi}, \quad A_t = A_t(r,\tilde r), \quad A_\varphi = A_\varphi(r,\tilde r)
\ee
Substituting this into the field equations, one obtains a set of nonlinear PDE's. The condensate is now a function of radius $\tilde r$. It vanishes at $\tilde r =0$ which  represents the center of the vortex, and approaches a constant at large $\tilde r$. This shows that there is no superconductivity inside the vortex. Since the Maxwell field is not dynamical on the boundary,  the magnetic field is not localized inside the vortex as one would expect. Instead, the authors of \cite{Montull:2009fe} work in finite volume and assume a uniform magnetic field on the boundary with total flux equal to $2\pi n$ as expected for the vortex.

\section{Recent developments}

So far, we have discussed planar black holes in the bulk. Spherical black holes have the same instability and describe spherical superconductors. One can add rotation to a spherical black hole, and the effect of this rotation on the dual superconductor has recently been studied \cite{Sonner:2009fk}.  It was found that for $T<T_c$ (the transition temperature at zero rotation), there is a critical value of the rotation which destroys the superconductivity in analogy to the critical  magnetic field. 

The bulk gravitational theory that we have studied was not derived from string theory. It was just the simplest model that could describe a dual superconductor, and we have seen that it works rather well.  It predicts that a charged condensate forms at low temperature (in a second order phase transition), that the  DC conductivity is infinite, and that there is a gap in the optical conductivity at low frequency -- all basic properties of superconductors. However, to go beyond this and have a more detailed microscopic understanding of the dual superconductor, one needs to embed this model in string theory. A linearized instability of the type needed to describe a holographic superconductor was found in a string compactification in \cite{Denef:2009tp}. This has now been extended to a full description by two different groups. Gauntlett et al. \cite{Gauntlett:2009dn} realized the $m^2 = -2$, $q=2$ model in M theory. In other words, they found a consistent truncation of eleven dimensional supergravity in which the four dimensional fields were just a metric, Maxwell field and charged scalar with this mass and charge\footnote{In general, there is an additional neutral scalar, but for purely electrically charged black holes, this field can be set to zero.}.
Gubser et al. \cite{Gubser:2009qm} realized the same model in one higher dimension (a five dimensional bulk which is dual to a $3+1$ dimensional superconductor) with     $m^2 = -3$, $q=2$ in type IIB string theory.
Both groups used Sasaki-Einstein compactifications with $U(1)$ symmetry, where the charged scalar is related to the size of the U(1) fibration. 

There are two main differences between the models obtained from string theory and our simple model. The first is that the kinetic term for the charged scalar involves nonminimal coupling. (This was investigated earlier by Franco et al. \cite{Franco:2009yz}.) The second difference is  that the scalar potentials $V(\psi)$ coming from supergravity are more complicated than the simple mass terms we have considered so far. This does not affect the onset of superconductivity: since the scalar field is small near $T_c$, the critical temperature is just determined by the mass term  in the potential. However, the low temperature limit is different. The potentials coming from supergravity  have more than one extremum. In this case, the zero temperature limit of the hairy black hole is a smooth domain wall in which the scalar rolls from one extremum in the UV to another extremum in the IR \cite{Gubser:2009gp}. The IR limit is another copy of $AdS$ with, in general, a different cosmological constant. 
These examples all have emergent conformal symmetry in the IR. 

The optical conductivity in these theories never show a pronounced gap at low temperature. Typically, $\Re \so = k\omega^\delta$  for small $\omega$ with a coefficient $k$ of order one. From our Schr\"odinger equation reformulation this is a bit surprising. Why isn't there an exponential suppression due to tunneling through the barrier? The answer is simply that the effective potential in the  Schr\"odinger equation never gets very large in these solutions since all fields are order one.

 Even though embedding in string theory makes it possible to study the dual field theory in more detail, this is still difficult in $2+1$ dimensions. It is somewhat easier in $3+1$ dimensions since the dual theories are related to ordinary gauge theory. 
 One particularly interesting aspect of the $3+1$ superconductor is that the condensate includes a term bilinear in the fermions, like a Cooper pair  \cite{Gubser:2009qm}.
 
  It was shown in \cite{Gubser:2008wv} that one can replace the Maxwell field and charged scalar in the bulk with an $SU(2)$ gauge field and still have a dual description of a superconductor. The main difference is the following: The symmetry of a superconductor refers to the energy gap above the Fermi surface. The holographic superconductors we have discussed here all have S-wave symmetry since they do not prefer a direction. With an $SU(2)$ gauge field in the bulk, one obtains a P-wave  superconductor\footnote{One cannot compute the energy gap directly. Indeed, as we have seen, there probably is not a strict gap. The interpretation as a P-wave superconductor comes from the fact that there is a vector order parameter and the conductivity is strongly anisotropic in a manor consistent with P-wave nodes on the Fermi surface.}. This leads to another way to realize holographic superconductors (in the probe limit) in string theory: one can use the $SU(2)$ gauge field realized on a pair of branes \cite{Ammon:2008fc,Peeters:2009sr}.
 
This is a rapidly evolving field and there have been several developments since the Milos summer school in September 2009, including: 
\begin{itemize}
\vskip .2cm
\item A key property of condensed matter systems is, of course,  the atomic lattice. This gives rise to the phonons in the BCS theory which cause the electrons to pair. Our simple model has translational symmetry and no sign of a lattice. However, some interesting ideas on how to include a lattice have been discussed by Kachru et al. \cite{Kachru:2009xf}.
\vskip .2cm
\item Some superconductors exhibit striped phases associated with charged density waves. It was shown in \cite{Nakamura:2009tf} that the Chern-Simons term in five dimensional supergravity can lead to an instability at nonzero momentum in which a spatially modulated condensate forms.  
\vskip .2cm
\item The response of fermions to the superconducting phase has been studied \cite{Chen:2009pt,Faulkner:2009am,Gubser:2009dt}.
 It was found that with a suitable coupling between the bulk fermions and scalar, there are stable quasiparticles with a  gap \cite{Faulkner:2009am}. 
\end{itemize}

\section{Conclusions and open problems}

 For  ninety years after its discovery in 1915, general relativity was viewed as a theory of gravity. It has proved very successful in describing a wide range of gravitational phenomena from the bending of light  to gravitational waves and black holes. But in the last few years we have seen that this same theory can describe other areas of physics as well, including superconductivity.  This is all due to the magic of  gauge/gravity duality. Although the full power of gauge/gravity duality relates a quantum theory of gravity (indeed string theory) in the bulk to a nongravitational theory on the boundary, we have worked in a large N limit in which the bulk theory is just classical general relativity. 
 This large N limit also explains how we can have spontaneous symmetry breaking in a $2+1$ dimensional field theory, in apparent contradiction to the Coleman-Mermin-Wagner theorem. The large N limit suppresses fluctuations in the fields.\footnote{This is not an issue for the $3+1$ dimensional superconductor,  which can be holographically described by the bulk gravitational theory (\ref{action}) in one higher dimension.}

It is natural to ask how surprised one should  be that general relativity can reproduce the basic properties of superconductors. After all, Weinberg \cite{Weinberg86} has shown that much of the phenomenology of superconductivity follows just from the spontaneous breaking of the $U(1)$ symmetry. Once we have found the instability that leads to charged scalar hair, doesn't everything else follow?  There are indications that something deeper is going on. For example,  order one dimensionless ratios can be computed and compared with experiment. In particular, the ratio $\omega_g/T_c \approx 8$ discussed in section 3.2,  is close to the observed value. This does not follow from  symmetry arguments alone.  

Since our bulk dual of a superconductor just involves gravity interacting with a Maxwell field and a charged scalar, there is a superficial similarity to a Landau-Ginzburg description. However, it is important to keep in mind two key differences. First, the low temperature instability must be put in by hand in the Landau-Ginzburg model, whereas it arises naturally in our gravitational description. Indeed, we have seen that there are two physically distinct instabilities which can trigger the phase transition. Second, the Landau-Ginzburg  model is only valid near the transition temperature, since it involves a power series in the order parameter $\varphi$. To go beyond $T\approx T_c$, one would need to specify an entire potential $V(\varphi)$. Initially, our bulk theory also had the freedom to add an arbitrary potential $V(\Psi)$. We chose just a mass term for simplicity. However, once one embeds the bulk theory into string theory, the potential is fixed and is no longer arbitrary.

\subsection{Open problems}

We close with a list of open problems\footnote{I thank Sean Hartnoll for suggesting some of these problems.}. They are roughly ordered in difficulty with the easier problems listed first. (Of course, this is my subjective impression. With the right approach, an apparently difficult problem may become easy!)

\begin{enumerate}
\vskip .2cm
\item In the probe limit below the critical temperature, there is an infinite discrete set of solutions for $\psi$ which are all regular on the horizon and satisfy the required boundary condition at infinity.  (When backreaction is included, there is only a finite number of such solutions.) They can be labeled by the number of nodes (zeros) they contain. It is widely believed that only the lowest solution with no nodes is stable. Is this true in the theory we have discussed where the scalar potential $V(\psi)$ just contains a mass term? Is it true if the theory contains scalar self interactions, like the potentials derived from string theory? 

\vskip .2cm
\item We have seen that an extreme Reissner-Nordstrom AdS black hole is unstable to forming neutral scalar hair if the scalar has mass close to the BF bound (\ref{eq:BF}). Under evolution, as the black hole is developing scalar hair, the total charge and mass at infinity are conserved. Since the scalar field can carry energy, but not charge, the final black hole (which need not be extremal) has the same charge, but less mass than the initial extremal limit. How is this consistent with the original black hole being extremal? What is the extremal limit in this Einstein-Maxwell-scalar theory? A related question is: what is the interpretation of this neutral scalar hair in the dual field theory?

\vskip .2cm
\item We studied magnetic fields in 2+1 dimensional superconductors in section 6. What happens if you add a magnetic field to   3+1 superconductors?  One immediate difference is that $B_c$ is no longer zero. Are the holographic superconductors still type II? The new magnetic brane solutions in \cite{D'Hoker:2009mm} may be useful.
\vskip .2cm
\item As discussed in section 6.2, a vortex solution has recently been found in the probe approximation. Find the vortex solution with backreaction. More generally, find the solution describing a lattice of vortices in the bulk. This would correspond to a magnetically charged black hole in which the magnetic flux is confined to ``cosmic strings" stretching from the horizon to infinity. To solve this problem, one must solve nonlinear partial differential equations.
\vskip .2cm
\item Recall that the symmetry of a superconductor refers to the energy gap on the Fermi surface. The holographic superconductors we have discussed here all have S-wave symmetry since they do not prefer a direction. P-wave holographic superconductors have been found \cite{Gubser:2008wv,Roberts:2008ns}. The high $T_c$ cuprates are known to have D-wave symmetry. Can one find a holographic superconductor with D-wave symmetry? It is natural to try to condense a charged spin two field. Interestingly enough, the newly discovered iron pnictides appear to have S-wave symmetry, but it is complicated by multiple Fermi surfaces \cite{chen:2008}.
\vskip .2cm
\item  It was discovered recently that there are materials in which the (DC) conductivity goes to zero for $T<T_c$ \cite{Vinokur:2008}. This is the opposite of a superconductor and is called a ``superinsulator". Can one find a holographic description of a superinsulator? It is tempting to use the fact that under electromagnetic duality in $AdS_4$, the conductivity transforms as $\sigma \rightarrow -1/\sigma$ \cite{Hartnoll:2007ip}. However one must start with a bulk action which is invariant under electromagnetic duality. The action (\ref{action}) is not, since it explicitly involves $A_\mu$.
\vskip .2cm
\item As we discussed in section 7,  it has recently been shown that the holographic superconductors can be derived from string theory, so in some cases the dual microscopic theory is known. Can one understand the pairing mechanism in these cases (if indeed ``pairing mechanism" is the right concept at strong coupling)? This remains one of the main open questions in understanding the high $T_c$ materials.
\vskip .2cm
\item As mentioned above, the large N limit is responsible for allowing spontaneous symmetry breaking in our 2+1 dimensional field theory. Can one show that away from this limit, massless fluctuations lead to infrared diverges which destroy the long range order?
\vskip .2cm
\item In gauge/gravity duality, one takes a large $N$ limit to justify using classical general relativity in the bulk. What is the analog of this large $N$ limit in condensed matter systems? In other words, what types of materials are likely to have a (tractable) dual gravitational gravitational description? (See \cite{polchinski2009} for a discussion of some of the issues.)
\vskip .2cm
\item  The high temperature cuprate superconductors satisfy a simple scaling law relating the superfluid density, the normal state (DC) conductivity and the critical temperature \cite{Homes:2004wv}. Can this be given a dual gravitational interpretation? 
\end{enumerate}

\begin{acknowledgement}
It is a pleasure to thank my collaborators, Sean Hartnoll, Chris Herzog, and Matt Roberts for teaching me many of the results described here. I also thank  Hartnoll and  Roberts for comments on these lecture notes. Finally, I  thank the organizers and participants of the 5th Aegean Summer
School, ``From Gravity to Thermal Gauge Theories: the AdS/CFT
Correspondence" for stimulating discussions. This work is supported in part by NSF grant number PHY-0855415.
\end{acknowledgement}

\end{document}